**Targeted Multispectral Filter Array Design for Endoscopic Cancer Detection in the Gastrointestinal Tract**

**Michaela Taylor-Williams[1,2], Ran Tao[1,2], Travis W Sawyer[3], Dale J Waterhouse[1,2,4], Jonghee Yoon[5], Sarah E Bohndiek[1,2]***

| | |
|---|---|
| 1 | Department of Physics, Cavendish Laboratory, University of Cambridge, JJ Thomson Avenue, Cambridge, CB3 0HE, UK |
| 2 | Cancer Research UK Cambridge Institute, University of Cambridge, Robinson Way, Cambridge, CB2 0RE, UK |
| 3 | Wyant College of Optical Sciences, University of Arizona, Tucson, USA |
| 4 | Now at Wellcome/EPRSC Centre for Interventional and Surgical Sciences (WEISS), University College London, Gower Street, London, WC1E 6BT, UK |
| 5 | Department of Physics, Ajou University, 16499, South Korea |

* Correspondence: Sarah E Bohndiek, Department of Physics, Cavendish Laboratory, University of Cambridge, JJ Thomson Avenue, Cambridge, CB3 0HE, UK and Cancer Research UK Cambridge Institute, University of Cambridge, Robinson Way, Cambridge, CB2 0RE, UK; seb53@cam.ac.uk; phone +441223 337267.

**Acknowledgements**

MT-W acknowledges the financial support of the General Sir John Monash Foundation, the Cambridge Trust International Scholarship, and the Cavendish Laboratory. RT acknowledges the financial support of the EPSRC Centre for Doctoral Training in Connected Electronic and Photonic Systems (EP/S022139/1), The Trinity Henry Barlow Scholarship, Cambridge Trust International Scholarship and the Herchel Smith Scholarship. TWS acknowledges support from the Winton Programme for the Physics of Sustainability and the United States Department of Defense (W81XWH-22-1-0211). SEB acknowledges support from the EPSRC (EP/R003599/1) and CRUK (C9545/A29580, C47594/A21102, C47594/A26851). We would like to thank George Gordon for his helpful input during manuscript preparation.

**Purpose:** Colour differences between healthy and diseased tissue in the gastrointestinal tract are detected visually by clinicians during white light endoscopy (WLE); however, the earliest signs of disease are often just a slightly different shade of pink compared to healthy tissue. Here, we propose to target alternative colours for imaging to improve contrast using custom multispectral filter arrays (MSFAs) that could be deployed in an endoscopic 'chip-on-tip' configuration.

**Methods:** Using an open-source toolbox, Opti-MSFA, we examined the optimal design of MSFAs for early cancer detection in the gastrointestinal tract. The toolbox was first extended to use additional classification models (k-Nearest Neighbour, Support Vector Machine, and Spectral Angle Mapper). Using input spectral data from published clinical trials examining the oesophagus and colon, we optimised the design of MSFAs with 3 to 9 different bands.

**Results:** We examined the variation of the spectral and spatial classification accuracy as a function of number of bands. The MSFA designs have high classification accuracies, suggesting that future implementation in endoscopy hardware could potentially enable improved early detection of disease in the gastrointestinal tract during routine screening and surveillance.

**Conclusion:** Optimal MSFA configurations can achieve similar classification accuracies as the full spectral data in an implementation that could be realised in far simpler hardware. The reduced number of spectral bands could enable future deployment of multispectral imaging in an endoscopic 'chip-on-tip' configuration.

**Keywords:** endoscopy, colonoscopy, oesophageal cancer, colon cancer, multispectral filter array, multispectral imaging.

# 1. Introduction

Multispectral imaging (MSI) is an emerging technique that holds promise in a range of biomedical applications, from monitoring of wound healing to enhancing contrast for early cancer in endoscopy [1–3]. MSI is based on the premise that tissues have their own spectrally unique reflectance fingerprint [4,5] arising from optical absorption and scattering processes, which are fundamentally altered by structural and biochemical changes that occur during disease progression. In cancer, for example, aberrant angiogenesis leads to neovascularisation that primarily alters optical absorption due to changes in haemoglobin (Hb) abundance and oxygenation levels. Furthermore, changes in cancer cell morphology, organelle distribution and size, alter tissue scattering properties [2,3,6–10]. Measurements of the spectral fingerprint of different tissue types have therefore been widely used to reveal the presence of cancer in excised tissue samples and *in situ* in patients [11–14].

Clinical white light endoscopy (WLE) and narrow band imaging (NBI) offer limited forms of MSI, targeting 3 colour (red, green, blue) vision and 2 colour (415 ± 10 nm and 540 ± 10 nm) haemoglobin absorption respectively. MSI incorporating a larger number of spectral bands is not yet widely used in clinical applications. Key challenges in clinical translation include the design of suitable instrumentation, as well as the development of appropriate expertise in operators and interpreters [2]. In terms of hardware, MSI systems typically require a trade-off between spectral, spatial and temporal resolution [5,15,16]. Broadly, spectral imaging can be implemented via four imaging system configurations: point-scanning 1D spectrometer, line-scanning 2D spectrometer, wavelength-scanning image sensor, or a snapshot imaging spectrometer [1,11]. The latter acquires the full 3D (x,y,wavelength) MSI datacube in a single acquisition, while the former options require scanning of either the spatial dimension(s) or the spectral dimension. These requirements can often lead to bulky and complex hardware, as well as offline data reconstruction. In endoscopy, real-time operation is required to ease operation and interpretation, as well as to account for patient movement during diagnostic procedures [11,17,18]. While snapshot MSI systems can achieve high temporal resolution, limited only by the camera frame rate, the low optical throughput and compromise between spatial and spectral resolution can degrade image quality compared to spatial-scanning systems.

To optimise the image quality in snapshot systems, one can target the spectral properties of the system to strategically sample incoming light at particular wavelengths known to be information rich in the target application of interest. The resulting spectral reflectance fingerprint of the given disease state can then be unmixed or classified accurately with fewer spectral samples [15]. Existing endoscopic systems use wavelength-scanning typically with a white light source and filter wheel for spectral sampling, however, this can become problematic in terms of temporal resolution and image co-registration as one increases the number of spectral samples. An alternative snapshot solution uses a multispectral filter array (MSFA) atop an imaging sensor [16,19,20]. Using absorptive or interference techniques, the MSFA filters the spectrum of the incoming light detected on a pixel-by-pixel basis, with a mosaic of filters deposited pixel-by-pixel across the sensor [3,21,22]. Following demosaicking and interpolation, the full image cube, including spatial and spectral information, is retrieved in a single snapshot [23]. Targeting the spectral properties of the MSFA to a given application thus maximises the spatial resolution [16,24,25] and is feasible from a manufacturing perspective [26,27], suggesting potential to address the aforementioned clinical unmet needs of MSI.

Here, we explore the potential for optimizing spectral sampling for early cancer detection in the gastrointestinal tract. We examine two published hyperspectral datasets available from prior endoscopy clinical trials, the first focused on detecting changes in tissue spectra in dysplasia and early cancer in the oesophagus and the second focused on measuring spectra from polyps and residual tissue post-resection in the colon. We expand the capability of the open-source Opti-MSFA toolbox to select the bandwidths and centre wavelengths for 3 – 9 filters [16]. Moreover, we then use the capability of the Opti-MSFA toolbox to tailor the spatial properties of an MSFA to sample these spectral bands according to a synthetic input hypercube with reference endmembers. We set parameters for the spectral properties of the filters using a merit function that represents the resulting error that would occur in sampling the spectra with the designed system [16]. Our results, derived using a range of merit functions, indicate the importance of end-to-end optimisation for customizing filters using appropriate merit functions when designing MSFAs. We demonstrate that customized spectral filters can be efficiently designed and optimised to detect gastrointestinal (GI) cancers using the Opti-MSFA toolbox based on the chosen datasets. These promising results suggest targeted design of MSFAs could provide similar MSI performance as full hyperspectral imaging, obviating the need for complex hyperspectral hardware in endoscopic applications.

## 2. Methods

The MSFA design exploited hyperspectral data collected during endoscopies of the oesophagus and colon in vivo (Section 2.1) [8,9], which was then analysed using the Opti-MSFA toolbox (Section 2.2) [16] to optimise the filters. Different classification techniques and unmixing functions were incorporated into the Opti-MSFA toolbox to develop the optimal MSFA for detecting cancers in these organs (Sections 2.3-2.5). Additionally, a merit function based on the unmixing of oxy- and deoxy-haemoglobin was used for comparison (Section 2.6). The code and synthetic datacubes used in the generation of this manuscript will be uploaded upon acceptance of the manuscript at: https://doi.org/10.17863/CAM.99953.

*2.1. Hyperspectral datasets of oesophageal and colon tissues*

Hyperspectral data (Figure 1) was collected in two prior clinical studies of hyperspectral endoscopy [8,9], the methods for which are briefly summarised below. Both data sets were collected using a 'babyscope' (PolyScope, Polydiagnost) that can be threaded through the accessory channel of a clinical gastroscope or colonoscope.

For the oesophageal study [8], the trial was reviewed by the Cambridgeshire Research Ethics Committee and was approved in March 2018 (18/NW/0134) and registered at ClinicalTrials.gov (NCT03388047). A broadband supercontinuum light source was used for illumination (SuperK COMPACT, NKT Photonics). The 10,000 fibrelet-imaging bundle (PD-PS-0095, PolyDiagnost) was imaged using an objective lens, and then the measured signal was split into two arms using a beam splitter. The split images were measured via a standard colour camera (Grasshopper 3.0, FLIR) and spectrometer (AvaSpec-ULS2048, Avantes; spectral range 200 to 1,100 nm, grating 300 lines/mm, slit size 50 mm,) to capture a structural image and averaged spectral information, respectively. Tissue spectra were collected in the oesophagus from three different tissue types, determined by histopathology: healthy squamous, non-dysplasia Barrett's oesophagus (NDBE), and neoplasia. These 715 spectra were collected from 15 different patients: 159 were from squamous regions; 320 from NDBE; and 236 from neoplastic regions.

For the colon study [9], the trial was reviewed by the OHSU Institutional Review Board (IRB18947) and registered at ClinicalTrials.gov (NCT04172493). The white light provided by the standard-of-care colonoscope was used for illumination (Olympus CF-H290). In the colon study, a 2D spectrograph (IsoPlane 160, Princeton Instruments; spectral range 400 to 800 nm, grating 150 lines/mm) was used with an electron-multiplied CCD camera (ProEM 1024, Princeton Instruments) to collect the spectral data using a line-scanning hyperspectral imaging method [28]. Tissue spectra were collected in the colon from three different tissue types determined by the colonoscopist: normal mucosa, polyp, and post-resection tissue (after polyp removal). In addition, spectral profiles of specular reflections were also captured. 14,065 spectra from ten different patients were collected; 5,269 spectra from normal regions; 1,045 from polyps, 7,745 post polyp resection, and 6 were a result of spectral reflectance.

Both spectral datasets used in the toolbox were downsampled using interpolation to have a spectral resolution of 1 nm, to simplify the computational process, and the wavelengths were restricted to between 470 and 720 nm, which eliminated spectral regions with high noise arising from insufficient illumination power.

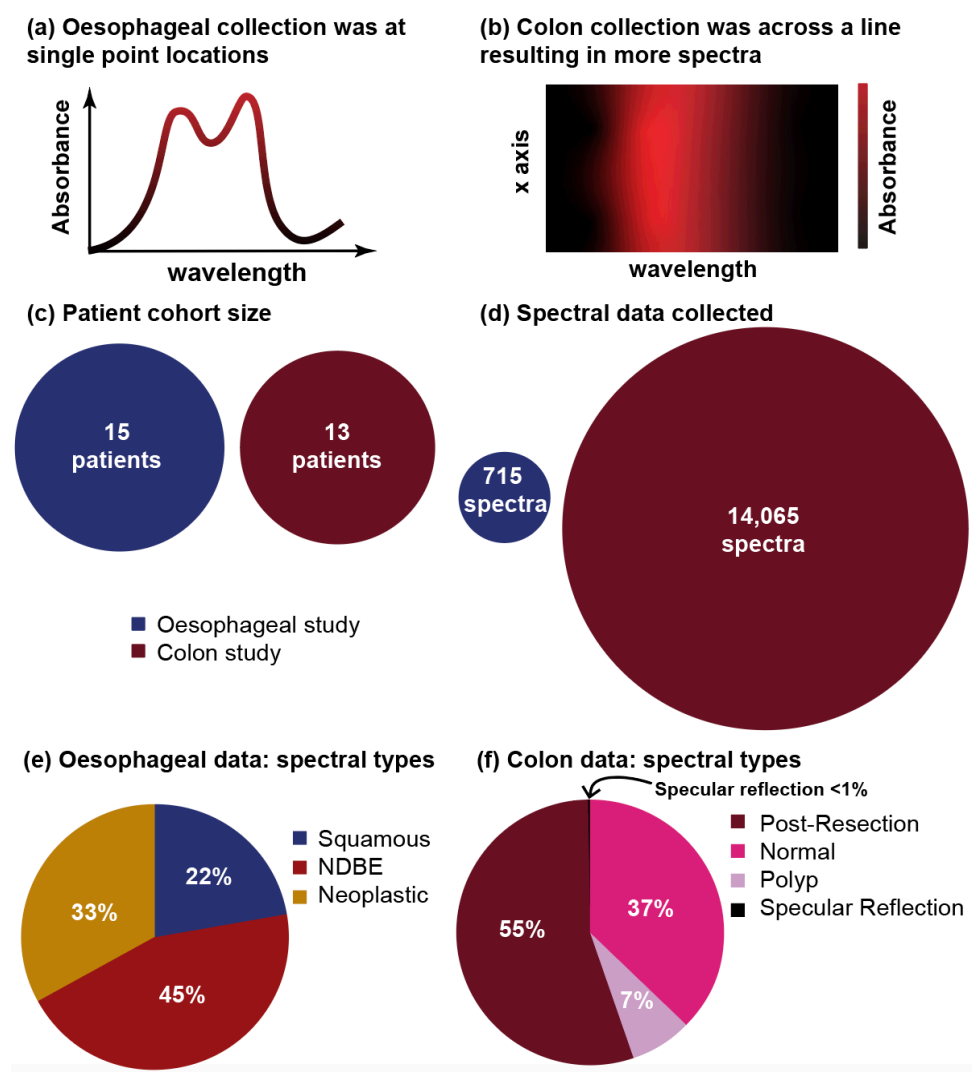

**Figure 1. An overview of the clinical datasets.** Illustration of the data collection method for the **(a)** oesophageal and **(b)** colon study. **(c)** Patient cohort size and **(d)** total spectra for each of the studies are compared to illustrate how the line scanning method results in significantly more data. **(e)** Oesophageal and **(f)** colon spectra are well-balanced between disease types.

*2.2. Hypercube generation*

To run the optimisation, synthetic hypercubes were created (Supplementary Figure 1) that were then input into the Opti-MSFA toolbox to optimise the filters. The hypercubes creation process differed slightly according to the different datasets (colon and oesophageal) and algorithms used for merit functions (if a training data split was required).

2.2.1 Splitting of data

For the hypercubes used with k-nearest neighbour (kNN) and support vector machines (SVM) classification models, the dataset was split in a 4:1 ratio for the training and testing (Supplementary Figure 1a,b). For spectral angle mapping and spectral unmixing, training is not required and the methods rely on identification of a reference spectrum, hence a training / testing split of the data was not required.

2.2.2 Oesophageal hypercube layout

For the oesophageal data, raw spectra were randomly subsampled (using a random number generator in Python) and allocated into regions in a 2x2 arrangement within an overall 80x80 shape (Supplementary Figure 1c,d). Raw spectra were selected randomly from either the testing set (kNN and SVM merit functions) or the full dataset (all other merit functions). The spatial arrangement of concentric circles for spectra from different disease types in the synthetic hypercube was designed to mirror the clinical situation where the detection of dysplasia within a background of non-dysplastic Barrett's oesophageal tissue is a key challenge; concentric circles containing dysplasia within NDBE within squamous tissue were created. The oesophageal hypercube was designed so that the ratio of the different tissue types in the hypercube approximated those of the underlying dataset, thus the resulting accuracies are weighted accordingly to address class imbalance.

2.2.3 Colon hypercube layout

For the colon data, 14,065 tissue spectra were randomly subsampled and placed in two distinct circular regions to represent the polyp and post-resection tissue surrounded by normal backgrounds, with overlapping specular reflections, again mirroring the clinical situation where polyps exist within a background of normal tissue. The colon hypercube created had a 96x96 shape with a spatial correlation of 2x2 regions, similar to the oesophageal hypercube. The shape was slightly larger than the oesophageal hypercube to allow for four spectral types instead of three, and the more complex spatial layout. For the unmixing hypercube, the same shape was used but size increased to 196x196 with a spatial correlation of 8x8 regions, since unmixing is more sensitive to discontinuities that occur at the boundaries due to the random generation of the hypercube.

2.2.4   Input of the hypercubes into the toolbox

The edges of the hypercubes after the classification were cropped, such that the oesophageal had a 70x70 shape and the colon had an 86x86 shape for classification or 186x186 shape for unmixing. This was to exclude misclassification due to the spatial demosaicking process at the edges of the MSFA sampling.

*2.3.   Opti-MSFA toolbox*

The open-source Python-based toolbox Opti-MSFA [16] was used to calculate optimal spectral and spatial filter properties. Input synthetic hypercubes were composed of endmember spectra (the spectral reflectance fingerprint of a given tissue type in this case) arranged in a suitable spatial pattern (Supplementary Figure 2a) as outlined in Section 2.2. Filter array simulation was then performed (Supplementary Figure 2b). Classification and unmixing-based merit functions were assessed using a gradient descent method to determine the optimal centre wavelength and bandwidth, defined as the full width at half maximum (FWHM). The spectral optimisation loop was run five times using different starting wavelengths to find five different candidate sets of spectral bands. The five different starting wavelength sets were selected by randomly sampling 10,000 candidate sets. 10,000 provided a sufficiently wide candidate set while not trying each option exhaustively and slowing down computation. The five filter sets with the highest classification accuracy in the initial tests were then refined using gradient descent. Allowed band centres were from 450 to 700 nm in 1 nm increments. Allowed FWHMs were from 10 to 30 nm in 2 nm increments to mirror the performance of fabricated MSFA. All possible spectral optimisation results in the five filter sets were similar, and the final spectral properties were taken as those with the highest classification accuracy. The filters with the highest classification accuracy were usually repeated within the five sets, especially when the number of filters was small (less than five bands), resulting in similar accuracies for the five sets. Non-matching filter sets typically implied that the gradient descent had converged on a local maximum rather than the desired global maximum.

Once the spectral bands were determined, they were then used to calculate the optimal spatial layout of the filters by an exhaustive search of all the possible spatial layouts. The possible spatial layouts are calculated by finding the Cartesian product of the different spectral bands, which effectively iterates all of the possible combinations of the filter bands:

$$\boldsymbol{B}(b_1, \ldots b_n)^{(i*j)} \tag{1}$$

where B are the n spectral bands $b_1$ to $b_n$, and i and j are the dimensions of the mosaic in question. The pattern that had the best classification accuracy following demosaicking (Supplementary Figure 2c, d) using a linear interpolation function was selected as the final optimised layout and the classification accuracy noted for comparison. Finally, the desired merit function was calculated (Supplementary Figure 2e). The entire process was repeated for n = 3 to 9 bands to determine optimised centre wavelengths, FWHMs and MSFA layouts for different MSI configurations.

*2.3    k Nearest Neighbours (kNN) classifier as a classification model*

The kNN algorithm is a supervised classification approach that classifies an image pixel by comparing its n-dimensional spectrum to the k closest n-dimensional endmember spectra in a labelled training set, where n is the number of spectral bands chosen for optimisation in Opti-MSFA. A given pixel is classified based on the consensus class of these k nearest neighbouring points [29]. A five-fold cross-validation was performed and the spectral region from 470 to 720 nm was assessed to determine the optimal k. The data was then shuffled to account for different combinations due to high patient variability in the underlying datasets. The process was repeated 20 times to account for the random shuffling of the data sets. Subsequently, the 20 repetitions were averaged, and it was found that k=5 gave the highest accuracy.

For optimisation of the centre wavelengths and FWHMs, the gradient descent algorithm was deployed. Endmember spectra were simulated by propagating the 'ground truth' spectra through the simulated filters using the individual spectral data that formed part of the testing data set. The function of merit was the accuracy of kNN classification for unmixing these simulated endmember spectra, where the kNN was trained on the training data set.

Once the optimum spectral filter set was determined, the spatial layout of filters was optimised. The different possible arrangements of the filters in a mosaic pattern were tested exhaustively in a 5-fold cross-validation process to find the layout with the highest classification accuracy. For each MSFA layout, imaging of the hypercubes was simulated by using the testing data set, which was then classified using the kNN algorithm. This was repeated four more times, so the spatial optimisation was found five times on five hypercubes and the classification accuracy was extracted for each variation. The overall classification accuracy was taken as the average classification accuracy across all five variations. The layout with the highest overall classification accuracy was chosen as the final optimised layout.

*2.4    Support Vector Machines (SVM) classifier as a classification model*

An alternate merit function using SVM classification was also implemented. SVM classifies data by defining hyperplanes to distinguish the data in n-dimensional space, where n is the number of bands in the filters [5,30]. The SVM classification parameters for the datasets were optimised in previous work by Waterhouse et al [31] where it was found that a radial basis function with C=1000 was best for performing SVM classification. Using the Sci-kit Learn toolbox available in Python 3, the Gaussian radial basis function kernel was used to define the hyperplane to approximate the different class distinctions. A mid-range value of C=1000, which effectively penalizes the function for misclassification, was again found to optimise classification when assessing the ground truth spectra using the individual spectra collected. The resulting classification accuracy was used as the merit function for SVM analysis of oesophageal and colon tissue.

SVM classification was implemented as a merit function. As for kNN, the classification accuracy of the filters was then tested on five simulated hypercubes in a 5-fold cross-validation process. The data set was split into a testing and training data set in a 1:4 ratio where the SVM algorithm was trained on training data and then tested on the hypercube made from the testing data (as outlined in Section 2.2). This was repeated a total of five times so the average classification accuracy of the five hypercubes was found and the layout with the highest classification accuracy was chosen as the optimal mosaic pattern.

*2.5    Spectral Angle Mapping (SAM) as a classification model*

In prior analysis of the colon dataset, it was noted that the collected data exhibited low variation in the spectral signatures of each class, which lends itself to using SAM classification [9]. With SAM, the spectra are represented as vectors and the angle between two vectors is calculated. Spectral classification is achieved by finding the reference spectrum that forms the smallest angle with the spectrum of interest by calculating the inner product using vector arithmetic [5,32]

$$\theta(\vec{t},\vec{r}) = \cos^{-1}\left(\frac{\sum_{i=1}^{n} t_i r_i}{\sqrt{(\sum_{i=1}^{n} t_i^2)} \cdot \sqrt{(\sum_{i=1}^{n} r_i^2)}}\right) \qquad (2)$$

where $\vec{t}$ and $\vec{r}$ are vectors that represent the spectrum of interest and the reference spectrum, respectively. For SAM, the classification accuracy is calculated as a ratio of the correctly classified spectra to the overall number of spectra classified. The spectral classification is assigned based on the tissue type with the lowest angle. SAM classification accuracy of 1 indicates that all testing data was correctly identified using SAM and 0 indicates that no testing data was identified correctly. SAM classification accuracy was tested as a merit function to optimise the spectral and spatial properties of the filters for 3 to 9 bands. Allowed band centres were from 470 to 700 nm in 1 nm increments. As for above, the classification accuracy of the filters was then trained and tested on five simulated hypercubes in a 5-fold cross-validation process. The resulting classification accuracy was used as the merit function for SAM analysis of oesophageal and colon tissue.

*2.6    Least squares spectral unmixing of haemoglobins*

In addition to classification methods, spectral data can also be subjected to linear spectral unmixing given prior knowledge of the main endmembers that contribute to the signal. In the context of biomedical tissue, oxy- ($HbO_2$) and deoxy- (Hb) haemoglobin are major optical absorbers that dominate optical absorption in the gastrointestinal tract. Scattering is also present and can be modelled. The reflectance spectra at a given point can be described by

$$R_r(\lambda) = c_{Hb}R_{Hb}(\lambda) + c_{HbO_2}R_{HbO_2}(\lambda) + c_{H_2O}R_{H_2O}(\lambda) + c_{scattering}R_{\text{scattering}}(\lambda) + c_{noise} \qquad (3)$$

where $R_r$, $R_{Hb}$, $R_{HbO_2}$, $R_{H_2O}$, and $R_{\text{scattering}}$ are the reflectance spectra of the pixel of interest *r*, for Hb, $HbO_2$, $H_2O$ and scattering, respectively [28]. The concentrations of deoxy- ($c_{Hb}$) and oxy- ($c_{HbO_2}$) haemoglobin molecules, water ($c_{H_2O}$), and scattering ($c_{scattering}$), noise present in the imaging system or tissue ($c_{noise}$) all contribute to the overall measured spectrum $R_r$. The reflectance spectra can be modelled as $R_t$ where $c'_{Hb}$, $c'_{HbO_2}$, and $c'_{scattering}$ are the estimated concentrations of deoxyhaemoglobin, oxyhaemoglobin, and scattering, respectively; in addition to an offset that accounts for noise and other errors, $c'_{offset}$:

$$R_t(\lambda) = c'_{Hb}R_{Hb}(\lambda) + c'_{HbO_2}R_{HbO_2}(\lambda) + c'_{H_2O}R'_{H_2O}(\lambda) + c'_{scattering}R_{\text{scattering}}(\lambda) + c'_{offset}$$
(4)

Fitting the above Equation 4 via a least-squares algorithm [33] optimises the concentration of deoxy- ($c'_{Hb}$) and oxy- ($c'_{HbO_2}$) haemoglobin, water ($c'_{H_2O}$), scattering ($c'_{scattering}$), and the offset ($c'_{offset}$) such that Equation 5 is minimized,

$$\sqrt{\sum_\lambda (R_t(\lambda) - R_r(\lambda))^2} \tag{5}$$

The Normalised Root Mean Square Error (NRMSE) function can be used to determine the unmixing accuracy while normalising using the mean HbO$_2$ and Hb values $\overline{c_{HbO_2}}$ and $\overline{c_{Hb}}$, respectively:

$$NRMSE = \frac{1}{p \cdot \overline{c_{Hb}} \cdot \overline{c_{HbO_2}}} \sqrt{\sum_p (c^*_{Hb} - c^{**}_{Hb})^2 + (c^*_{HbO_2} - c^{**}_{HbO_2})^2} \tag{6}$$

Where *p* is the number of pixels; in the spectral optimisation it is the full data set, while in the spatial optimisation it is the demosaicked hypercube. $c^*_{HbO_2}$ and $c^*_{Hb}$ are the concentrations of HbO$_2$ and Hb calculated using the full spectra while $c^{**}_{HbO_2}$ and $c^{**}_{Hb}$ are the concentrations of HbO$_2$ and Hb calculated using the reduced spectra of interest at the selected centre wavelengths and bandwidths.

The oxygen saturation (sO$_2$) of the different tissue types was calculated as a ratio of the concentration of HbO$_2$ to the sum of HbO$_2$ and Hb,

$$sO_2 = \frac{c_{HbO_2}}{c_{Hb} + c_{HbO_2}} \tag{7}$$

$$v_{blood} = c_{Hb} + c_{HbO_2} \tag{8}$$

The relative fraction of blood *v$_{blood}$* was also calculated and plotted against the sO$_2$ for the different tissue types.

*2.7 Classification accuracy when using the full spectra*

To assess the compromise in classification accuracy when using a limited number of bands for MSI we compared to classification of the tissue using hyperspectral imaging. A hyperspectral filter with centre wavelengths from 470 and 720 nm in 1 nm steps was applied to the spectra with bandwidths (FWHM) of 1 to 30 nm in 1 nm steps (Supplementary Figure 3). When machine learning was used for classification for the kNN and SVM techniques, it was done in a 5-fold method, in line with the classification techniques used on the hypercubes. The resulting classification accuracy for each tissue type and classification method was calculated following spectral band optimisation in the Opti-MSFA toolbox; spatial optimisation was not relevant in this case.

## 3. Results

*3.1. Optimised filters designed for classification of oesophagus and colon*

The optimal 3 to 9 band MSFA arrangements for oesophageal and colon datasets are determined according to classification accuracies, where high values reflect high performance when using kNN, SVM, and SAM, while low values of NMRSE for unmixing $HbO_2$ and Hb are preferred. A direct comparison across both datasets for all merit functions showing optimised filter arrays and output images is available in the Supplementary Information (Supplementary Tables 1-3). For reference, we calculated the classification accuracies also for a 250-band sample, considering the performance that could be achieved with a full hyperspectral imaging system (Supplementary Figure 3, Table 1), showing overall a greater separability of the spectra in the colon dataset than the oesophageal.

**Table 1.** Full hyperspectral classification accuracy of oesophageal and colon tissue depending on classification type.

| Classification model | Maximum Classification Accuracy | |
|---|---|---|
| | Oesophagus | Colon |
| kNN | 0.848 | 0.999 |
| SVM | 0.811 | 0.997 |
| SAM | 0.245 | 0.995 |

When considering the oesophageal dataset (Tables 2, 3), the spectral bands highlighted in the 3-band case using the kNN (475±24nm, 573±22nm and 617±16nm) are then generally represented in all filter sets with higher numbers of bands (Figure 2). When adding a fourth band, a far-red region (703±16nm) is added and again remains present through 6 and 9 band examples. The optimised outputs show consistent sampling of the haemoglobin absorption region in the 525 to 575 nm range, usually with additional bands flanking outside of this region, one towards the blue and one towards the red. A strong representation of filters in the green is also seen in the SVM results, however, in this case the red and far-red spectral bands are only represented in the 7-9 band cases. For both kNN and SVM (Figure 2a,c), the classification accuracy after spectral optimisation is largely unaffected by adding more spectral bands (Figure 2b,d), suggesting the blue-green target filters in the 3 band case are already sufficiently information rich, consistent with prior use of haemoglobin targeted narrow-band imaging approaches clinically. Spatial optimisation leads to degradation in performance as would be expected (Supplementary Figure 4), since more spectral samples lead to poorer spatial sampling. SAM appears to make the most use of the red and far-red spectral bands (Figure 2e,f) but shows poor classification accuracy in the oesophageal dataset, most likely because subtle overall changes in signal intensity are important for classifying the oesophageal tissue spectra and these are not accounted for in the SAM dataset.

A similar trend is seen for the colon dataset in terms of both the information rich spectral features and the classification performance for kNN (Figure 3a,b) and SVM (Figure 3c,d). For the colon data set, the central region of sampling is slightly red-shifted to 550 to 600 nm range, with further sampling in the red common, particularly for SAM (Figure 3e,f). All methods show perfect or near-perfect classification accuracy (Tables 4, 5) following the spectral optimisation, owing to the more separable spectral features in the dataset, which leads to less reliance on global intensity changes. To examine this further, a 95% confidence interval was calculated by finding the average classification accuracy or unmixing error for each patient and then performing a t-test on these averages; these are shown in plots of classification accuracy and unmixing error. Inter-patient variation is also lower than in the oesophageal data as there is a smaller number of biological replicates (patients). We would expect a performance degradation with an expanded dataset with a more representative sampling of inter-patient variation. After spatial optimisation, the performance degrades somewhat but remains above 95% for 3- and 4-band solutions; adding more filters does not lead to improvement in performance and in fact reduces classification accuracy, suggesting this is adding noise. The results indicate that it is not necessary to go beyond 3 or 4 spectral band MSI for the classification of these datasets and to do so may detriment the outcome.

**Table 2**. Overall classification accuracy of oesophageal tissue using kNN, SVM, and SAM after spectral optimisation.

| Classification model | Number of Filters | | | | | | |
|:---:|:---:|:---:|:---:|:---:|:---:|:---:|:---:|
| | 3 | 4 | 5 | 6 | 7 | 8 | 9 |
| kNN | 0.839 | 0.848 | 0.850 | 0.860 | 0.851 | 0.855 | 0.850 |
| SVM | 0.836 | 0.837 | 0.850 | 0.844 | 0.841 | 0.841 | 0.830 |
| SAM | 0.476 | 0.473 | 0.477 | 0.417 | 0.359 | 0.351 | 0.343 |

**Table 3.** Overall classification accuracy of oesophageal tissue using kNN, SVM, and SAM after both spectral and spatial optimisation.

| Classification model | Number of Filters | | | | | | |
|:---:|:---:|:---:|:---:|:---:|:---:|:---:|:---:|
| | 3 | 4 | 5 | 6 | 7 | 8 | 9 |
| kNN | 0.803 | 0.804 | 0.795 | 0.800 | 0.778 | 0.801 | 0.824 |
| SVM | 0.797 | 0.817 | 0.789 | 0.756 | 0.792 | 0.757 | 0.756 |
| SAM | 0.335 | 0.329 | 0.333 | 0.360 | 0.235 | 0.324 | 0.240 |

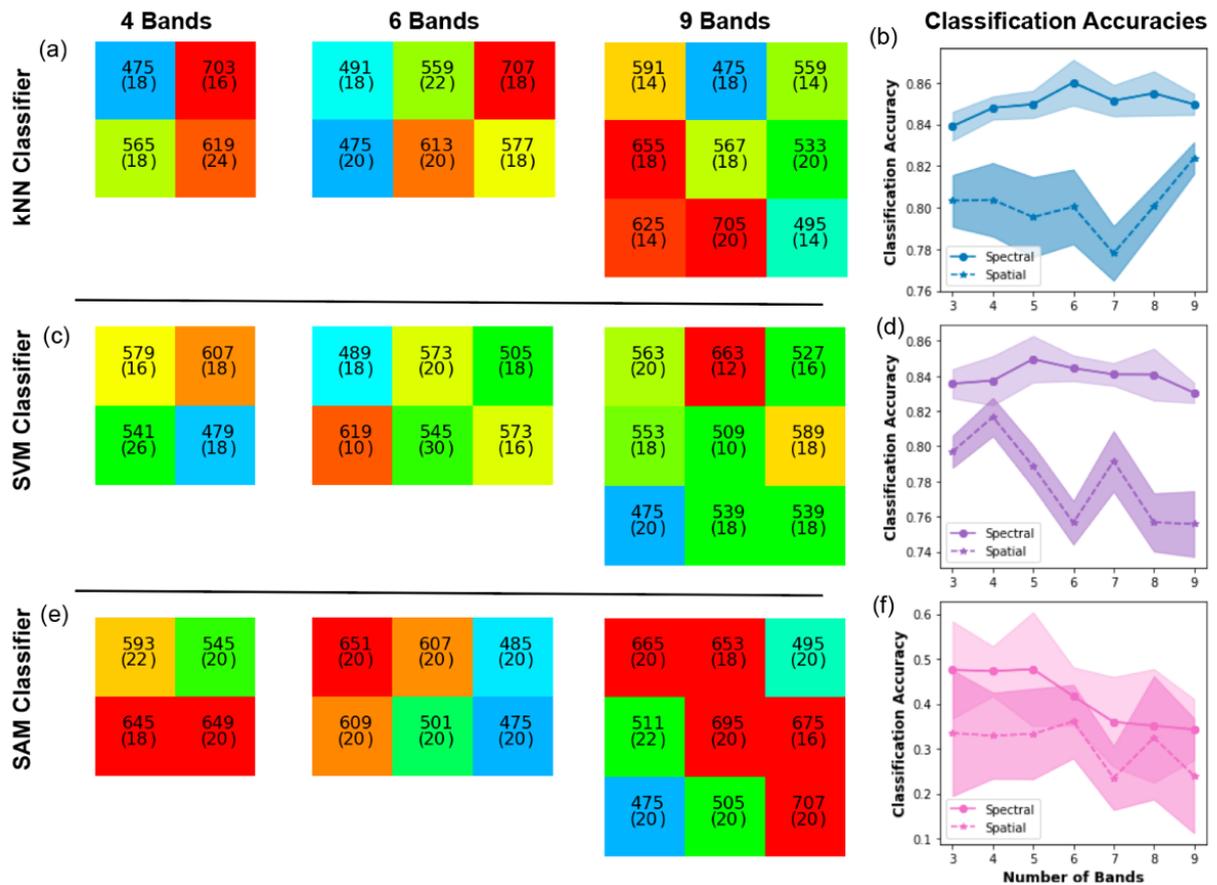

**Figure 2. Optimal band arrangements and classification accuracy for the oesophageal tissue types**. Using **(a,b)** kNN classification, **(c,d)** SVM classification, and **(e,f)** SAM classification. **(a,c,e)** The optimised spatial layouts of the filters are shown, with each coloured box indicating the spectral properties, including the centre wavelength and FWHM bandwidth (given in parenthesis). The spectral properties are shown in nanometres. The colours of the individual filters approximate what the human eye would see at these wavelengths. **(b,d,f)** Classification accuracies after spectral and spatial optimisation and the associated 95% confidence intervals.

**Table 4.** Overall classification accuracy of colon tissue using kNN, SVM, and SAM after spectral optimisation.

| Classification model | Number of Filters | | | | | | |
|:---:|:---:|:---:|:---:|:---:|:---:|:---:|:---:|
| | 3 | 4 | 5 | 6 | 7 | 8 | 9 |
| kNN | 1.0000 | 1.0000 | 1.0000 | 1.0000 | 1.0000 | 1.0000 | 1.0000 |
| SVM | 1.0000 | 1.0000 | 1.0000 | 1.0000 | 1.0000 | 1.0000 | 1.0000 |
| SAM | 0.9993 | 0.9995 | 0.9997 | 0.9996 | 0.9999 | 0.9997 | 0.9999 |

**Table 5.** Overall classification accuracy of colon tissue using kNN, SVM, and SAM, after spectral and spatial optimisation.

| Classification model | Number of Filters | | | | | | |
|:---:|:---:|:---:|:---:|:---:|:---:|:---:|:---:|
| | 3 | 4 | 5 | 6 | 7 | 8 | 9 |
| kNN | 0.976 | 0.963 | 0.963 | 0.954 | 0.950 | 0.946 | 0.946 |
| SVM | 0.961 | 0.955 | 0.944 | 0.941 | 0.931 | 0.931 | 0.926 |
| SAM | 0.968 | 0.964 | 0.961 | 0.955 | 0.952 | 0.952 | 0.946 |

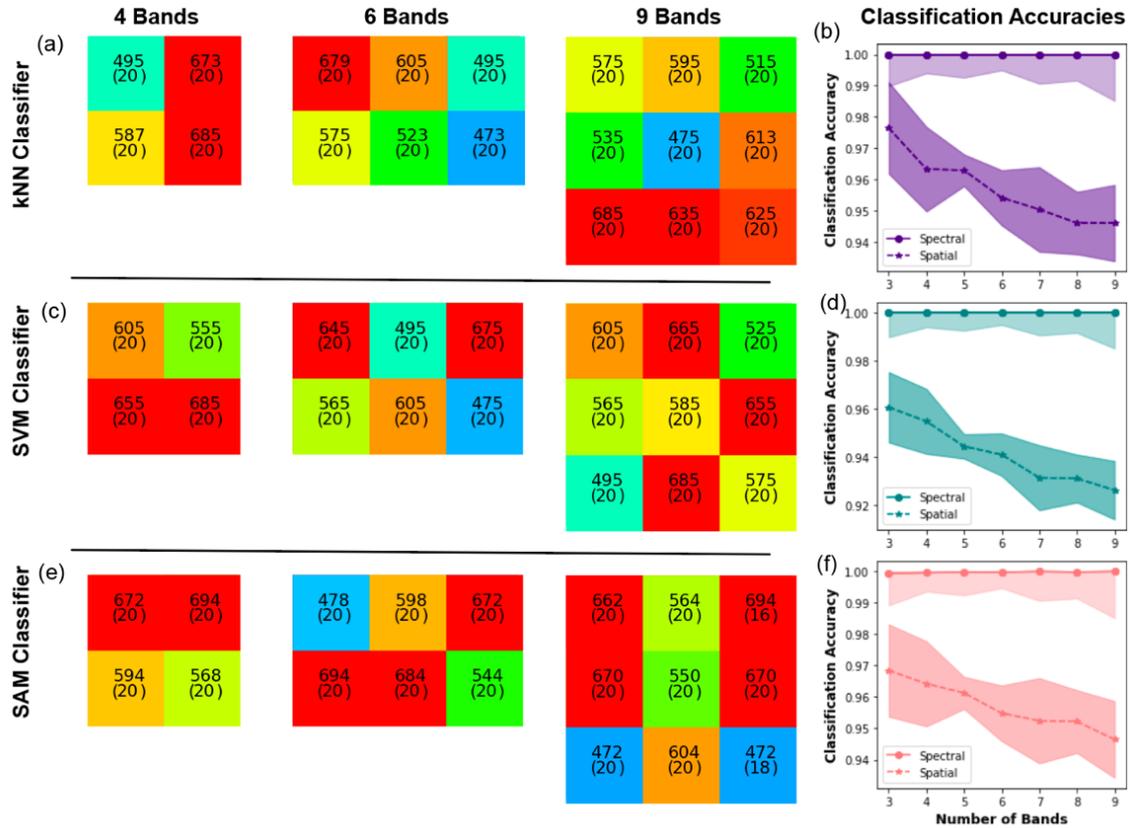

**Figure 3. Optimal band arrangements and classification accuracy for the colon tissue types.** Using **(a,b)** kNN classification, **(c,d)** SVM classification, and **(e,f)** SAM classification. **(a,c,e)** The spatial arrangement of the filters is shown as per Figure 4. **(b,d,f)** Classification accuracies after spectral and spatial optimisation and the associated 95% confidence intervals.

*3.2 Optimisation based on linear spectral unmixing for haemoglobin content and oxygenation*

We next examined the potential to apply linear spectral unmixing to the recorded spectra for the assessment of physiological parameters of haemoglobin content and oxygenation and using the NRMSE to determine the goodness of fit with different numbers of filters. Consistent with prior findings,[8] the oesophageal data set was not clearly separable based on these parameters (Figure 4a), however, the colon dataset, newly examined using this approach, was highly separable (Figure 4b). Not only was it possible to resolve post resection bleeding from polyps, it was also possible to distinguish polyps from normal tissue using these metrics. We therefore proceeded to undertake spectral band optimisation only for the colon data set.

The goodness of fit for the measured reflectance spectra to reference spectra for constituent oxy- and deoxy-haemoglobin components, along with water and background scattering, was first examined across each colon tissue type. Overall, the fit performance was high for normal tissue, ranging from $r^2$ = 0.82 to 0.99 (Supplementary Table 4), but as expected showed a greater variance in the diseased states, ranging from $r^2$ = 0.54 to 0.99 in polyp tissue (Supplementary Table 5) and $r^2$ = 0.48 to 0.99 in post-resection tissue (Supplementary Table 6). The majority of the fits are in the upper $r^2$ range and the few lower performing fits are due to noise in the spectra (Supplementary Figures 5-7). Undertaking the optimisation process on these spectra (Figure 4c) again shows the need for strong sampling in the green, but now flanked by a bluer and redder band (510±10nm, 558±18nm and 618±30nm). Interestingly the red band in this case is rather broad compared to the other merit functions, perhaps linked to the fit to the background scattering term. The unmixing error declines with increasing number of bands for spectral unmixing, indicating that it is easier to fit the reference spectra with increased sampling (Figure 4d, Table 5).

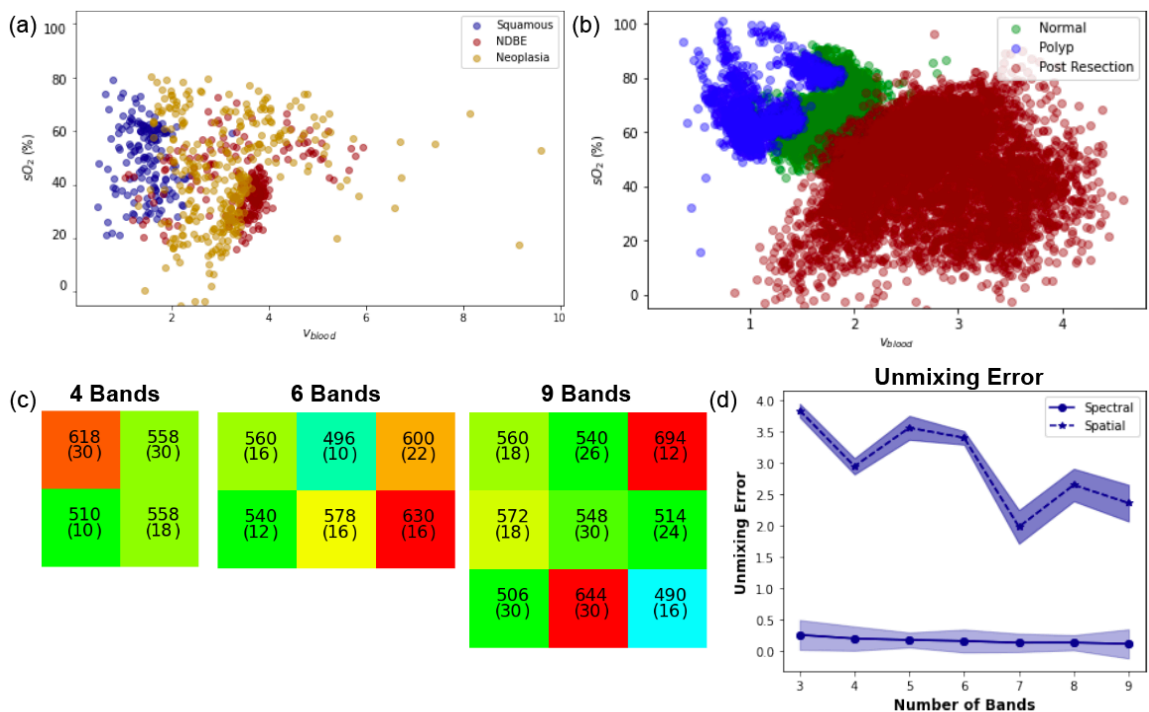

**Figure 4. Unmixing of oxygen saturation (SO$_2$) and blood volume ($v_{blood}$). (a)** Poor differentiation for tissue types in the oesophagus is seen based on oxygenation, consistent with prior publication. Conversely, better discrimination is seen **(b)** for the various tissue types in the colon, hence the colon dataset only was subject to MSFA optimisation based on linear unmixing. **(c)** The filters optimised using unmixing on colon tissue types. **(d)** Unmixing error after spectral and after spatial optimisation and the associated 95% confidence intervals.

**Table 5.** Spectral unmixing NRMSE of colon tissue after spectral and spatial optimisation.

| | Number of Filters | | | | | | |
|---|---|---|---|---|---|---|---|
| Classification type | 3 | 4 | 5 | 6 | 7 | 8 | 9 |
| After spectral optimisation | 0.250 | 0.192 | 0.170 | 0.153 | 0.125 | 0.127 | 0.107 |
| After spatial optimisation | 3.832 | 2.942 | 3.556 | 3.395 | 1.975 | 2.644 | 2.354 |

*3.3. Filter classification illustrated on synthetic hypercubes*

Overall, the output results from Opti-MSFA indicate that both kNN and SVM classification had similar accuracies in oesophageal tissue, reflecting the higher inter-patient variation in the oesophageal dataset.[8] The classification accuracy of the colon tissue was much higher as the colon spectra are more separable, as found in prior examination using SAM. [9] We next explored how the optimal choice of spectral bands would translate into imaging performance. In the oesophageal dataset (Figure 5), the reconstruction of the concentric circles of different tissue types is well reflected in the kNN and SVM, with some 'spillover' at the boundaries between regions. A higher level of misclassification of the normal tissue is seen using SVM. The SAM approach gives a poor performance, with a circular structure only barely apparent in the 9-band case and even then showing an inversion of the disease types spatially. The performance of the MSFA-based approach clearly decreases as spatial resolution is lost in the 6-band and 9-band mosaic patterns.

In the colon dataset (Figure 6), classification performance is again degraded at the edges of features, with poorer edge definition seen with an increasing number of filters. Interestingly the edges of the specular reflections are consistently misclassified first to polyp and then to normal tissue with distance from the centre. This would be important clinically since specular reflections are image artefacts and could be misleading if classified as polyp tissue. In the case of unmixing (Figure 7), spectral resolution is more important and outweighs the deterioration due to spatial resolution (Supplementary Table 7).

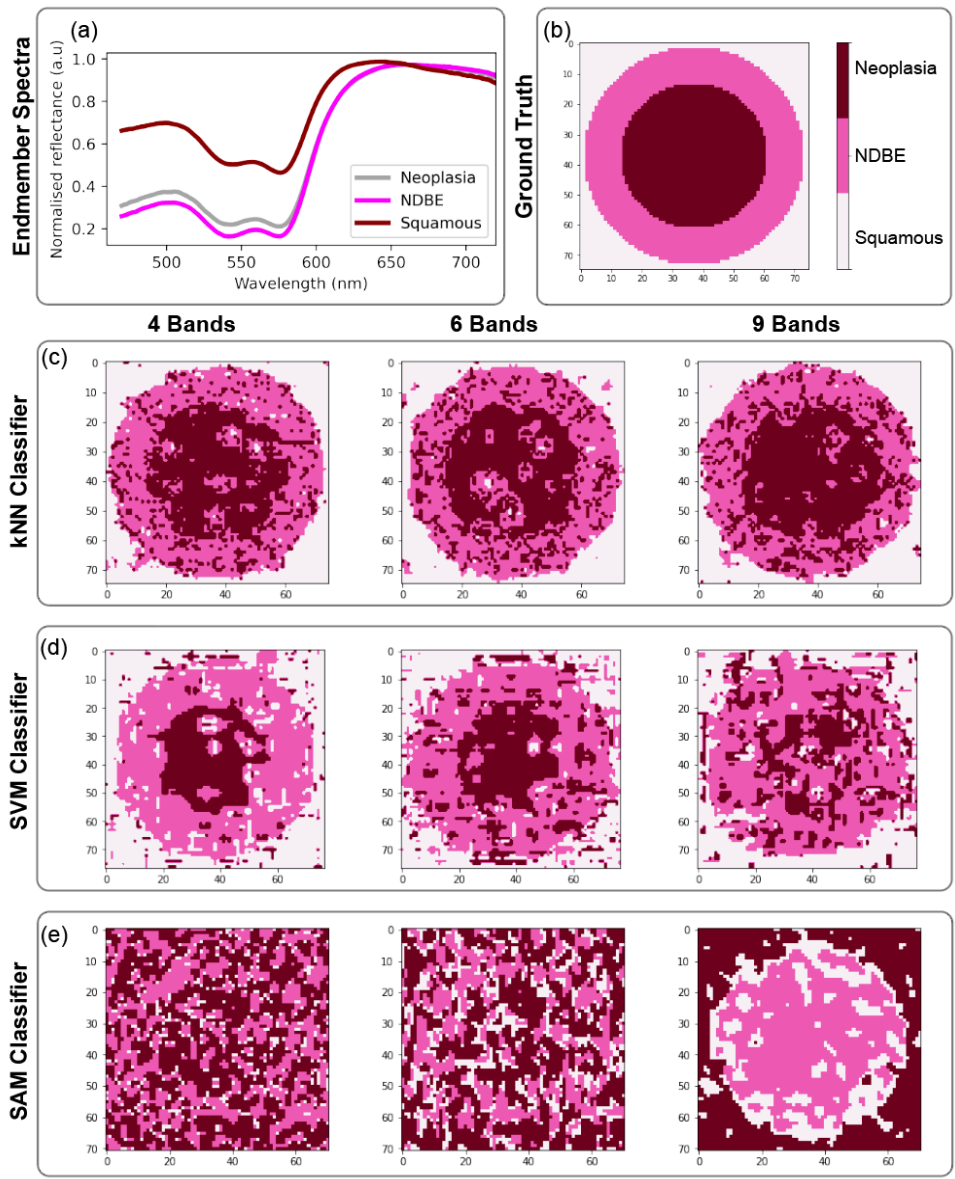

**Figure 5. Oesophageal classification imaging outputs. (a)** The spectral endmembers and **(b)** the ground truth of the synthetic hypercubes used in the optimisation process. The resulting synthetic hypercube image classifications using the different MSFA designs are shown for **(c)** kNN, **(d)** SVM, and **(e)** SAM.

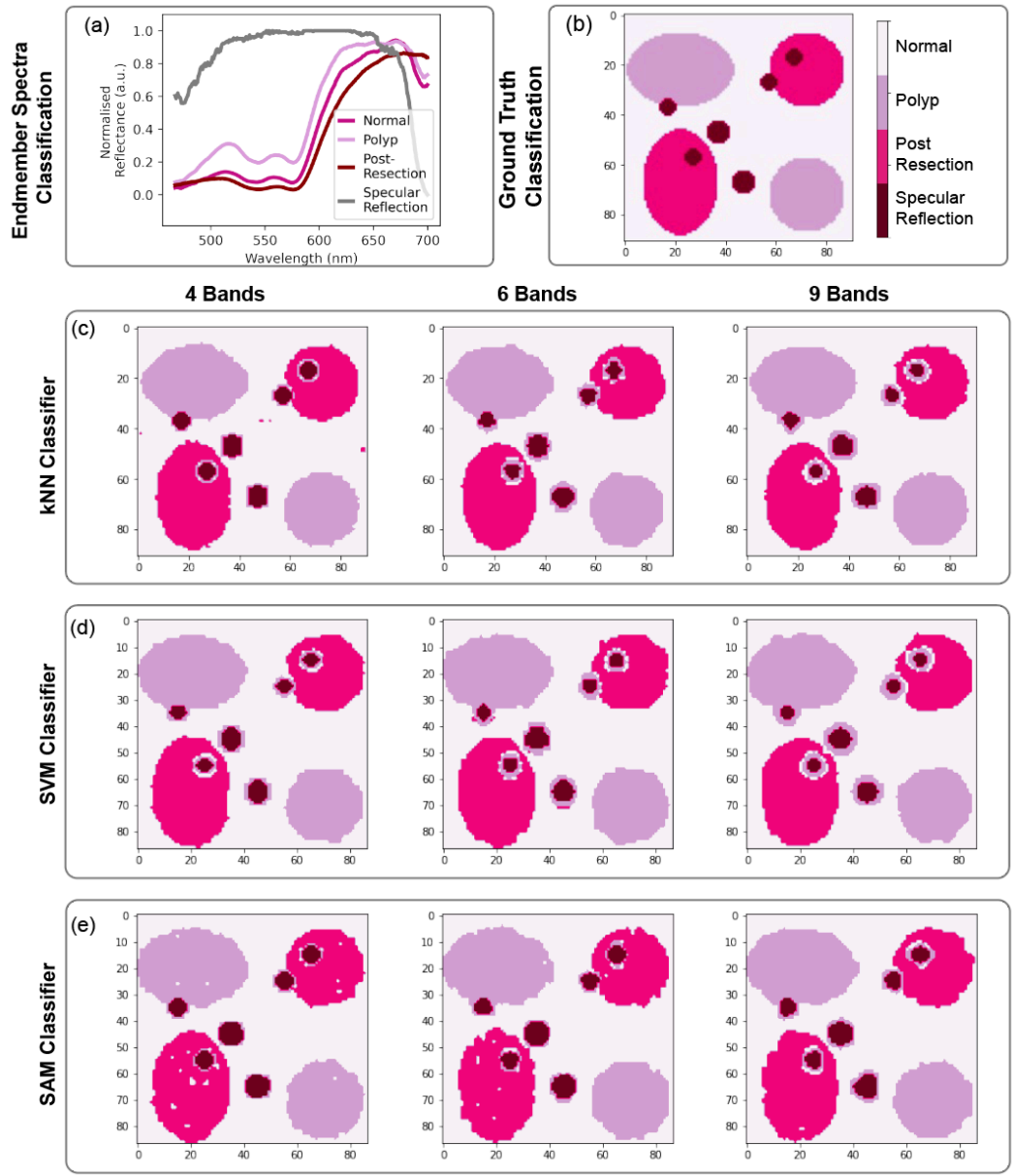

**Figure 6. Colon classification imaging outputs. (a)** The spectral endmembers and **(b)** the ground truth of the synthetic hypercubes used in the optimisation process. The resulting synthetic hypercube image classifications using the different MSFA designs are shown for **(c)** kNN, **(d)** SVM, and **(e)** SAM.

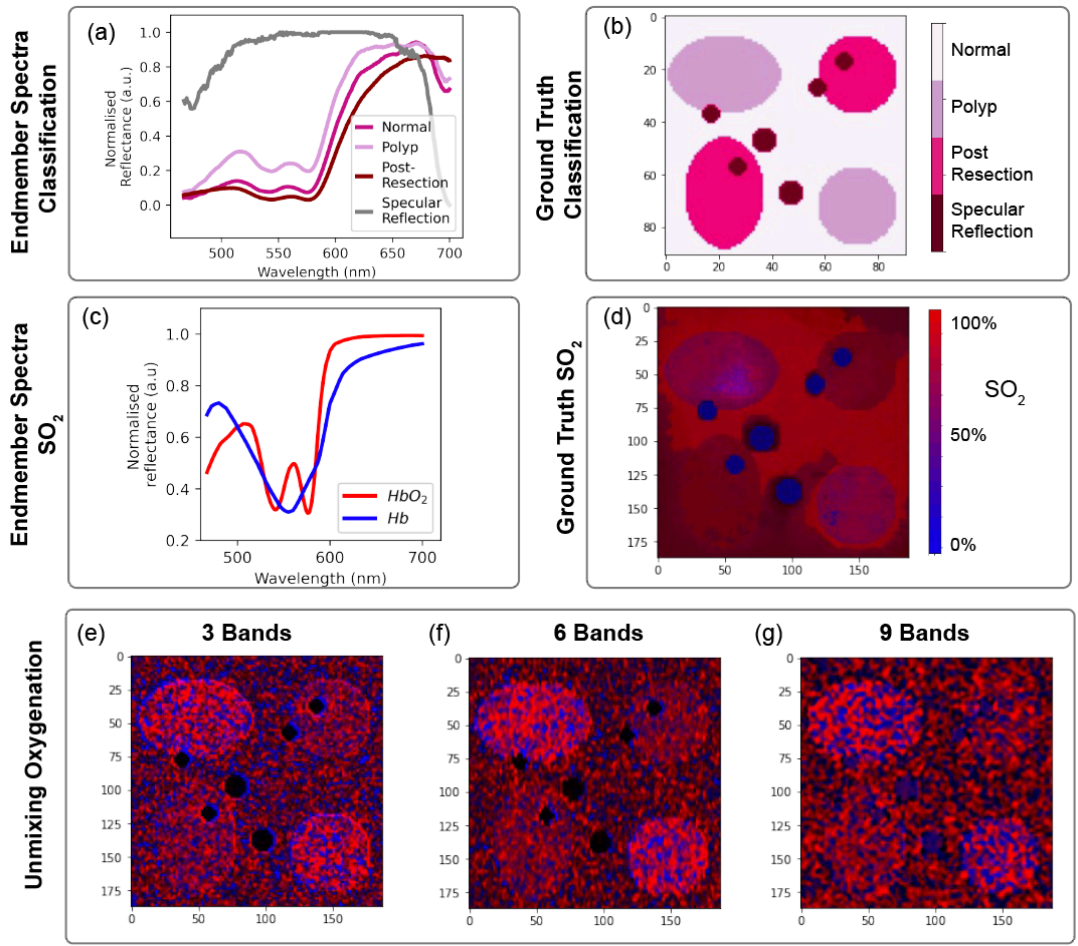

**Figure 7. Colon spectral unmixing imaging outputs. (a)** The spectral endmembers and **(b)** ground truth of the synthetic hypercubes used in the classification optimisation process are illustrated. **(c)** The corresponding spectral endmembers and **(d)** ground truth in the unmixing of $SO_2$ are illustrated. **(e)** The unmixing $SO_2$ abundance maps for the MSFA optimisation are also shown.

## 4. Discussion

Here, we applied an open-source software toolbox Opti-MSFA to analyse published datasets with the goal to identify the potential for multispectral imaging to enhance contrast during endoscopy for early cancer detection. Our analysis of the optimal wavelengths and bandwidths for discrimination of early cancerous lesions in the gastrointestinal tract shows promise, even with only 3 or 4 targeted wavebands. The classification accuracies achieved for a highly restricted bandset were very similar to those obtained for the full hyperspectral dataset, indicating that MSI using information rich spectral bands presents a good strategy to trading spectral and spatial information.

Testing the MSI approach across different merit functions and datasets revealed some interesting insights. In both datasets, there was typically a central spectral sampling range in which one or more spectral band appeared, flanked by a bluer and a redder spectral band. Bands present in the 3-band case then appeared similarly as an optimisation output as the number of bands increased. The precise location of these bands varied slightly between the oesophageal and colon data sets, with the central sampling range being 525 to 575 nm in the former and 550 to 600 nm in the latter.

The bandwidth of the targeted spectral bands was generally quite broad in all cases, on the order of 20 nm, likely due to the presence of relatively smoothly varying spectral features. It is likely that the wider bandwidth balances the trade-off of improved spectral resolution that occurs with a small bandwidth and the additional noise that occurs due to the reduced amount of light that reaches the image sensor since everything outside of this band is rejected. This was not the case when the unmixing accuracy of $HbO_2$ and Hb was calculated, with spectral bandwidths as small as 10 nm and as large as 30 nm, the minimum and maximum bandwidths allowed, indicating that spectral sensitivity is important when unmixing $HbO_2$ and Hb. The use of wider bands is advantageous from the perspective of optical system design for future applications, requiring less stringent design specifications for optical components, but may not be tolerated for applications where unmixing for physiological metrics is preferred over disease classification.

Finally, the trade-off between addition of spectral information and reduction in spatial resolution must be considered. The classification accuracy after spatial optimisation for both datasets generally decreased as the number of bands in the filters increased, except for spectral unmixing where the NRMSE improved with more bands. It is worth noting that spectral optimisation assumes a complete hypercube can be measured, whereas the spatial optimisation results in a sparse hypercube due to the multispectral filter array. With the classification models, as the number of bands increases, there is a higher chance of adding a less information rich, noisier, band and the lost spatial information then outweighs the spatial information. Some variation is seen according to the optimisation metric used. While SAM is the simplest and least computationally intensive method, it relies on distinct spectral- rather than intensity-based changes, which means that it shows a poor performance in the oesophageal dataset where the spectral changes between tissue types are more heavily reliant on intensity changes. Comparing kNN and SVM, kNN can better discern the edges of the neoplastic or polyp regions against the background. The decrease in classification accuracy with number of bands is seen most significantly in the SVM classifiers, where increasing the number of bands from 3 to 9 reduces accuracy by 0.08, while the kNN classifier only decreases by 0.007. These findings reinforce the importance of choosing an appropriate classifier for the dataset under investigation, considering spatial and spectral information and noise characteristics. Further work is necessary to fully understand the best choice of merit function for a given set of data and how it will influence the performance of the resulting MSFA.

While the analysis performed here is promising for the use of targeted spectral bands in the gastrointestinal tract, it is based on hypothetical hypercubes using a relatively restricted set of published spectra. Further analysis should evaluate how inter- and intra-patient variation, along with noise variations in imaging systems, might affect the optimal filter arrays. The testing and training datasets here were split before training, but since there is inter-patient variation, this could lead to over-classification. To assess this, the 95% confidence interval was calculated on the average classification accuracies of different patients. The highest classification accuracies occurred in the colon dataset, in agreement with the classification in previous work that used SAM methods very and showed very good separation of the spectra of the different tissue types. Further work could assess how these classification accuracies vary between patients and consider confounding factors that may influence the measurements, such as age or genetics. To assess the impact of the spatial variation in true hyperspectral data, future work should acquire full hyperspectral imaging datacubes from endoscopy *in situ* to optimise MSFAs. Other sources of error that would be encountered in a clinical setting, such as patient motion, should also be examined. Nevertheless, the results shown here demonstrate that the complex trade-offs involved with spectral imaging can be balanced by tuning the MSFAs using different merit functions to focus on classification or unmixing.

5. **Conclusion**

Customised MSFAs with a relatively small number of spectral bands could be applied to enhance contrast for early cancer detection in the GI tract.

**Statements and Declarations**

SEB, TWS and DJW are co-inventors on a patent relating to spectral band optimization. The other authors declare no conflicts of interest.

Author Contributions: Concept and design: MT-W, TWS, SEB; Data curation and Methodology: MT-W, DJW, RT, JY; Investigation and analysis: MT-W, RT; Visualization: MT-W; Writing–original draft: MT-W; Writing–review and editing: MT-W, RT, TWS, DJW, JY, SEB.

Copyright: For the purpose of open access, the authors have applied a Creative Commons Attribution (CC BY) licence to any Author Accepted Manuscript version arising.

## 6. References

1. Clancy, N.T.; Jones, G.; Maier-Hein, L.; Elson, D.S.; Stoyanov, D. Surgical Spectral Imaging. *Med. Image Anal.* **2020**, *63*, 101699, doi:10.1016/j.media.2020.101699.

2. Waterhouse, D.J.; Fitzpatrick, C.R.M.; Pogue, B.W.; O'Connor, J.P.B.; Bohndiek, S.E. A Roadmap for the Clinical Implementation of Optical-Imaging Biomarkers. *Nat. Biomed. Eng.* **2019**, *3*, 339–353, doi:10.1038/s41551-019-0392-5.

3. Taylor-Williams, M.; Spicer, G.; Bale, G.; Bohndiek, S.E. Noninvasive Hemoglobin Sensing and Imaging: Optical Tools for Disease Diagnosis. *J. Biomed. Opt.* **2022**, *27*, 080901, doi:10.1117/1.JBO.27.8.080901.

4. Li, Q.; He, X.; Wang, Y.; Liu, H.; Xu, D.; Guo, F. Review of Spectral Imaging Technology in Biomedical Engineering: Achievements and Challenges. *J. Biomed. Opt.* **2013**, *18*, 100901, doi:10.1117/1.jbo.18.10.100901.

5. Lu, G.; Fei, B. Medical Hyperspectral Imaging: A Review. *J. Biomed. Opt.* **2014**, *19*, 010901, doi:10.1117/1.jbo.19.1.010901.

6. Roy, S.; Kumaravel, S.; Sharma, A.; Duran, C.L.; Bayless, K.J.; Chakraborty, S. Hypoxic Tumor Microenvironment: Implications for Cancer Therapy. *Exp. Biol. Med.* **2020**, *245*, 1073–1086, doi:10.1177/1535370220934038.

7.  Saito, T.; Yamaguchi, H. Optical Imaging of Hemoglobin Oxygen Saturation Using a Small Number of Spectral Images for Endoscopic Imaging. *J. Biomed. Opt.* **2015**, *20(12)*, 126011, doi:10.1117/1.JBO.20.12.126011.

8.  Waterhouse, D.J.; Januszewicz, W.; Ali, S.; Fitzgerald, R.C.; Di Pietro, M.; Bohndiek, S.E. Dataset for: Spectral Endoscopy Enhances Contrast for Neoplasia in Surveillance of Barrett's Esophagus. *Cancer Res.* **2021**, *81*, 3415–3425, doi:10.17863/CAM.61809.

9.  Yoon, J.; Joseph, J.; Waterhouse, D.J.; Borzy, C.; Siemens, K.; Diamond, S.; Tsikitis, V.L.; Bohndiek, S.E. Dataset for: First Experience in Clinical Application of Hyperspectral Endoscopy for Evaluation of Colonic Polyps. *J. Biophotonics* **2021**, 1–9, doi:0.17863/CAM.61810.

10. Maktabi, M.; Köhler, H.; Ivanova, M.; Jansen-Winkeln, B.; Takoh, J.; Niebisch, S.; Rabe, S.M.; Neumuth, T.; Gockel, I.; Chalopin, C. Tissue Classification of Oncologic Esophageal Resectates Based on Hyperspectral Data. *Int. J. Comput. Assist. Radiol. Surg.* **2019**, *14*, 1651–1661, doi:10.1007/s11548-019-02016-x.

11. Hagen, N.; Kudenov, M.W. Review of Snapshot Spectral Imaging Technologies. *Opt. Eng.* **2013**, *52*, 090901, doi:10.1117/1.oe.52.9.090901.

12. Aboughaleb, I.H.; Aref, M.H.; El-Sharkawy, Y.H. Hyperspectral Imaging for Diagnosis and Detection of Ex-Vivo Breast Cancer. *Photodiagnosis Photodyn. Ther.* **2020**, *31*, 101922, doi:10.1016/j.pdpdt.2020.101922.

13. Martinez, B.; Leon, R.; Fabelo, H.; Ortega, S.; Piñeiro, J.F.; Szolna, A.; Hernandez, M.; Espino, C.; O'shanahan, A.J.; Carrera, D.; et al. Most Relevant Spectral Bands Identification for Brain Cancer Detection Using Hyperspectral Imaging. *Sensors (Switzerland)* **2019**, *19*, doi:10.3390/s19245481.

14. Panasyuk, S. V.; Yang, S.; Faller, D. V.; Ngo, D.; Lew, R.A.; Freeman, J.E.; Rogers, A.E. Medical Hyperspectral Imaging to Facilitate Residual Tumor Identification during Surgery. *Cancer Biol. Ther.* **2007**, *6*, 439–446, doi:10.4161/cbt.6.3.4018.

15. Sawyer, T.W.; Williams, C.; Bohndiek, S.E. Spectral Band Selection and Tolerancing for Multispectral Filter Arrays. *Front. Opt. - Proc. Front. Opt. + Laser Sci. APS/DLS* **2019**, 3–5, doi:10.1364/FIO.2019.JW4A.126.

16. Sawyer, T.W.; Taylor-Williams, M.; Tao, R.; Xis, R.; Williams, C.; Bohndiek, S.E.; Xia, R.; Williams, C.; Bohndiek, S.E. Opti-MSFA: A Toolbox for Generalized Design and Optimization of Multispectral Filter Arrays. *Opt. Express* **2022**, *30*, 7591, doi:10.1364/oe.446767.

17. Waterhouse, D.J.; Fitzpatrick, C.R.M.; di Pietro, M.; Bohndiek, S.E. Emerging Optical Methods for Endoscopic Surveillance of Barrett's Oesophagus. *Lancet Gastroenterol. Hepatol.* **2018**, *3*, 349–362, doi:10.1016/S2468-1253(18)30030-X.

18. Taylor-Williams, M.; Mead, S.; Sawyer, T.W.; Hacker, L.; Williams, C.; Berks, M.; Murray, A.; Bohndiek, S.E. Multispectral Imaging of Nailfold Capillaries Using Light-Emitting Diode Illumination. *J. Biomed. Opt.* **2022**, *27*, 126002, doi:10.1117/1.JBO.27.12.126002.

19. Wu, R.; Li, Y.; Xie, X.; Lin, Z. Optimized Multi-Spectral Filter Arrays for Spectral Reconstruction. *Sensors (Switzerland)* **2019**, *19*, doi:10.3390/s19132905.

20. Yang, Z.; Albrow-Owen, T.; Cai, W.; Hasan, T. Miniaturization of Optical Spectrometers. *Science (80-. ).* **2021**, *371*, doi:10.1126/science.abe0722.

21. Daqiqeh Rezaei, S.; Dong, Z.; You En Chan, J.; Trisno, J.; Ng, R.J.H.; Ruan, Q.; Qiu, C.W.; Mortensen, N.A.; Yang, J.K.W. Nanophotonic Structural Colors. *ACS Photonics* **2021**, *8*, 18–33, doi:10.1021/acsphotonics.0c00947.

22. Shambat, G.; Mirotznik, M.S.; Euliss, G.; Smolski, V.O.; Johnshon, E.G.; Athale, R. Photonic Crystal Filters for Multi-Band Optical Filtering on a Monolithic Substrate. *J. Nanophotonics* **2009**, *3*, 031506, doi:10.1117/1.3110223.

23. Li, P.; Asad, M.; Horgan, C.; MacCormac, O.; Shapey, J.; Vercauteren, T. Spatial Gradient Consistency for Unsupervised Learning of Hyperspectral Demosaicking: Application to Surgical Imaging. *Int. J. Comput. Assist. Radiol. Surg.* **2023**, doi:10.1007/s11548-023-02865-7.

24. Wu, H.M.; Lee, T.A.; Ko, P.L.; Liao, W.H.; Hsieh, T.H.; Tung, Y.C. Widefield Frequency Domain Fluorescence Lifetime Imaging Microscopy (FD-FLIM) for Accurate Measurement of Oxygen Gradients within Microfluidic Devices. *Analyst* **2019**, *144*, 3494–3504, doi:10.1039/c9an00143c.

25. Nouri, D.; Lucas, Y.; Treuillet, S. Hyperspectral Interventional Imaging for Enhanced

Tissue Visualization and Discrimination Combining Band Selection Methods. *Int. J. Comput. Assist. Radiol. Surg.* **2016**, *11*, 2185–2197, doi:10.1007/s11548-016-1449-5.

26. Miyamichi, A.; Ono, A.; Kamehama, H.; Kagawa, K.; Yasutomi, K.; Kawahito, S. Multi-Band Plasmonic Color Filters for Visible-to-near-Infrared Image Sensors. *Opt. Express* **2018**, *26*, 25178, doi:10.1364/oe.26.025178.

27. Taylor-Williams, M.; Cousins, R.B.; Williams, C.; Bohndiek, S.E.; Mellor, C.J.; Gordon, G.S.D. Spectrally Tailored "hyperpixel" Filter Arrays for Imaging of Chemical Compositions. *Proc. SPIE BiOS* **2022**, *11954*, doi:10.1117/12.2606917.

28. Yoon, J.; Joseph, J.; Waterhouse, D.J.; Luthman, A.S.; Gordon, G.S.D.; di Pietro, M.; Januszewicz, W.; Fitzgerald, R.C.; Bohndiek, S.E. A Clinically Translatable Hyperspectral Endoscopy (HySE) System for Imaging the Gastrointestinal Tract. *Nat. Commun.* **2019**, *10*, 1–13, doi:10.1038/s41467-019-09484-4.

29. Zhang, Z. Introduction to Machine Learning: K-Nearest Neighbors. *Ann. Transl. Med.* **2016**, *4*, 1–7, doi:10.21037/atm.2016.03.37.

30. Bahari, N.I.S.; Ahmad, A.; Aboobaider, B.M. Application of Support Vector Machine for Classification of Multispectral Data. *IOP Conf. Ser. Earth Environ. Sci.* **2014**, *20*, doi:10.1088/1755-1315/20/1/012038.

31. Waterhouse, D.J.; Januszewicz, W.; Ali, S.; Fitzgerald, R.C.; di Pietro, M.; Bohndiek, S.E. Spectral Endoscopy Enhances Contrast for Neoplasia in Surveillance of Barrett's Esophagus. *Cancer Res.* **2021**, *81*, 3415–3425, doi:10.1158/0008-5472.CAN-21-0474.

32. Ran, L.; Zhang, Y.; Wei, W.; Zhang, Q. A Hyperspectral Image Classification Framework with Spatial Pixel Pair Features. *Sensors (Switzerland)* **2017**, *17*, 1–20, doi:10.3390/s17102421.

33. Wei, J.; Wang, X. An Overview on Linear Unmixing of Hyperspectral Data. *Math. Probl. Eng.* **2020**, *2020*, doi:10.1155/2020/3735403.

**Supplementary Information for: Targeted Multispectral Filter Array Design for Endoscopic Cancer Detection in the Gastrointestinal Tract**

Michaela Taylor-Williams[1,2], Ran Tao[1,2], Travis W Sawyer[3], Dale Waterhouse[4], Jonghee Yoon[5], Sarah E Bohndiek[1,2]*

1   Department of Physics, Cavendish Laboratory, University of Cambridge, JJ Thomson Avenue, Cambridge, CB3 0HE, UK
2   Cancer Research UK Cambridge Institute, University of Cambridge, Robinson Way, Cambridge, CB2 0RE, UK
3   Wyant College of Optical Sciences, University of Arizona, Tucson, USA
4   Wellcome/EPRSC Centre for Interventional and Surgical Sciences (WEISS), University College London, Gower Street, London, WC1E 6BT, UK
5   Department of Physics, Ajou University, 16499, South Korea

* Correspondence: Sarah E Bohndiek, Department of Physics, Cavendish Laboratory, University of Cambridge, JJ Thomson Avenue, Cambridge, CB3 0HE, UK and Cancer Research UK Cambridge Institute, University of Cambridge, Robinson Way, Cambridge, CB2 0RE, UK; seb53@cam.ac.uk; phone +441223 337267.

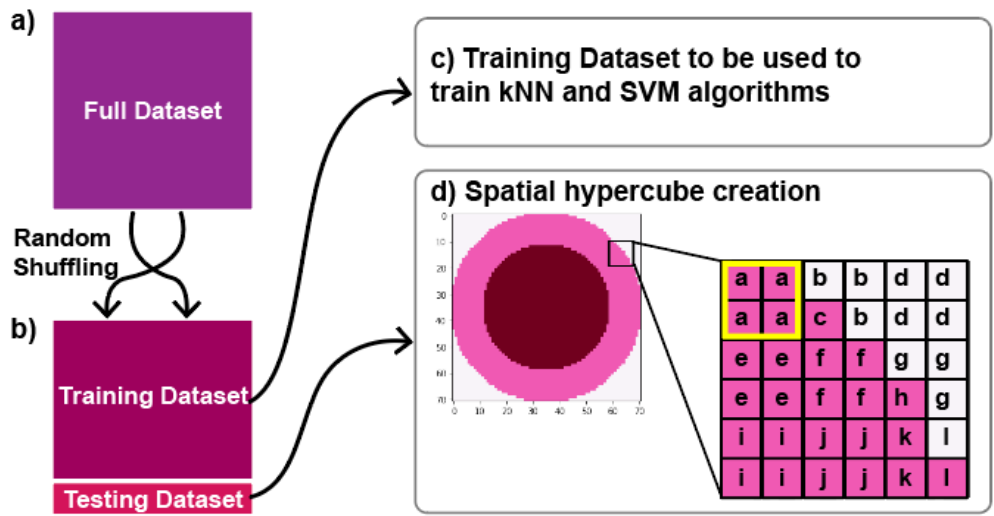

**Supplementary Figure 1. The process of synthetic hypercube creation. (a)** Hypercubes use with the full data set or a randomly shuffled data subset split in testing and training datasets (for kNN and SVM classification). **(b)** The training data set is used to train the algorithm of interest and **(c)** the testing dataset is used to make a spatial hypercube, such that 2x2 regions have the same spectra as shown by the letters illustrating different spectral data randomly selected from the testing data set.

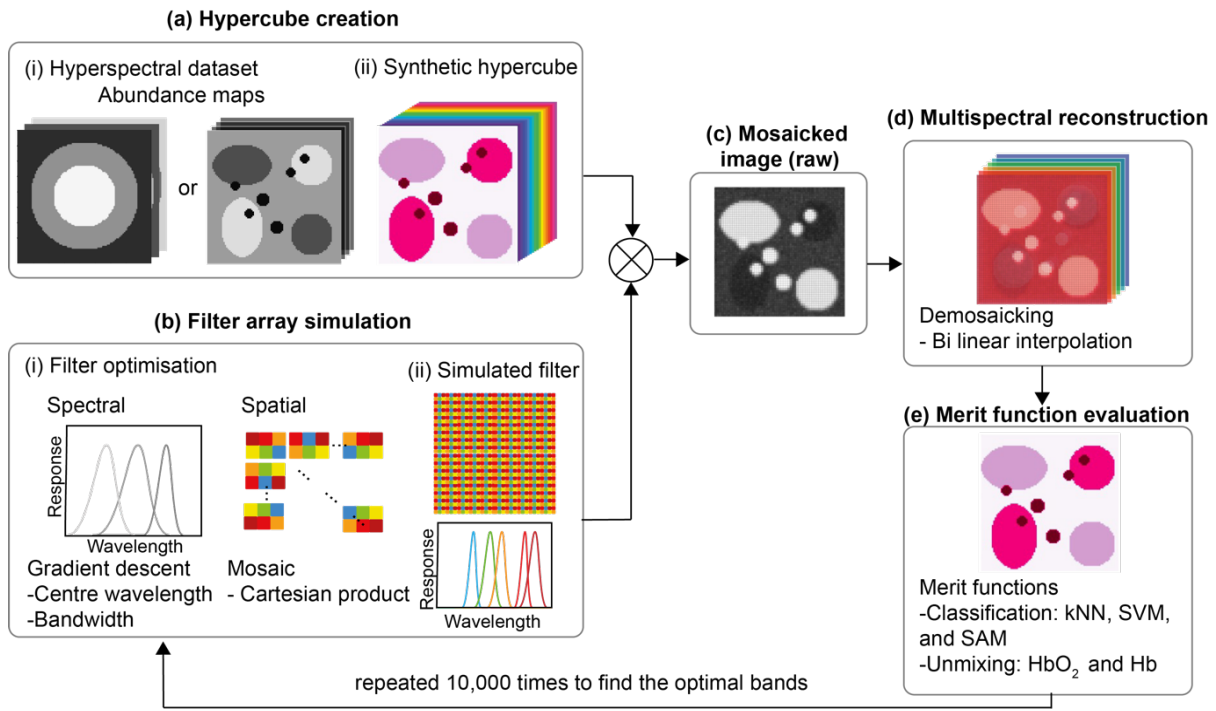

**Supplementary Figure 2**. **The Opti-MSFA toolbox recreates the classification and unmixing process after imaging tissue using the MSFA** [15] **(a)** A hypercube is created by inputting **(i)** spectral datasets and their abundance maps to create **(ii)** a synthetic hypercube. Based on this hypercube and **(b)** a simulated filter array **(c)** a raw mosaicked image is then simulated. **(d)** The mosaicked image is then reconstructed using bilinear interpolation. **(e)** Different classification and unmixing techniques were used for endmember reconstruction, and these were used as merit functions to optimize the **(b, i)** spectral and spatial properties of possible filters that are then **(b, ii)** simulated and tested.

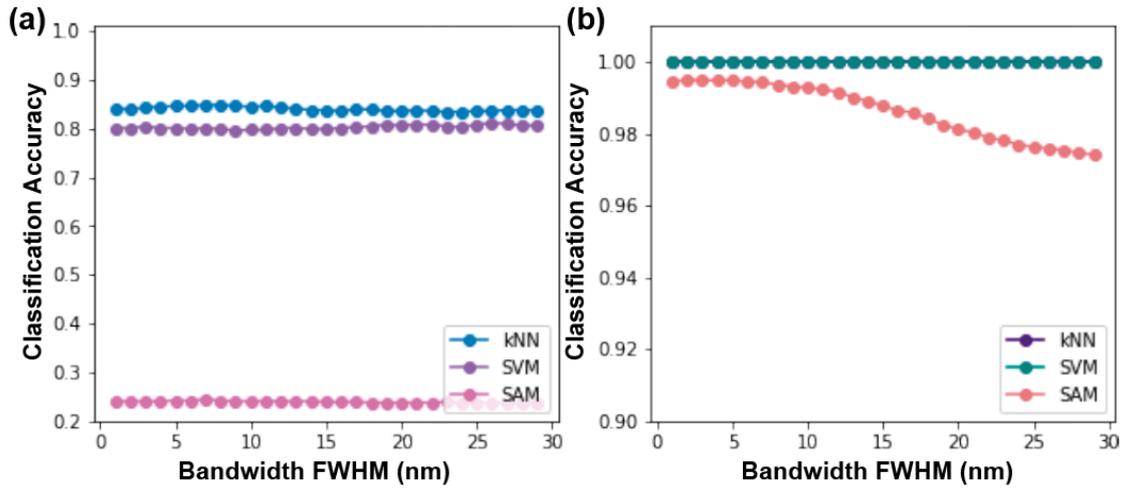

**Supplementary Figure 3. Classification of (a) oesophageal and (b) colon tissue with 250 filters with varying bandwidths.** Classification accuracies were calculated for 250 bands (from 470 to 720 nm). The bandwidths of the filters were then assessed from 1 to 30 nm. The classification accuracy of the colon using kNN equivalent to the SVM, so the data points are not visible.

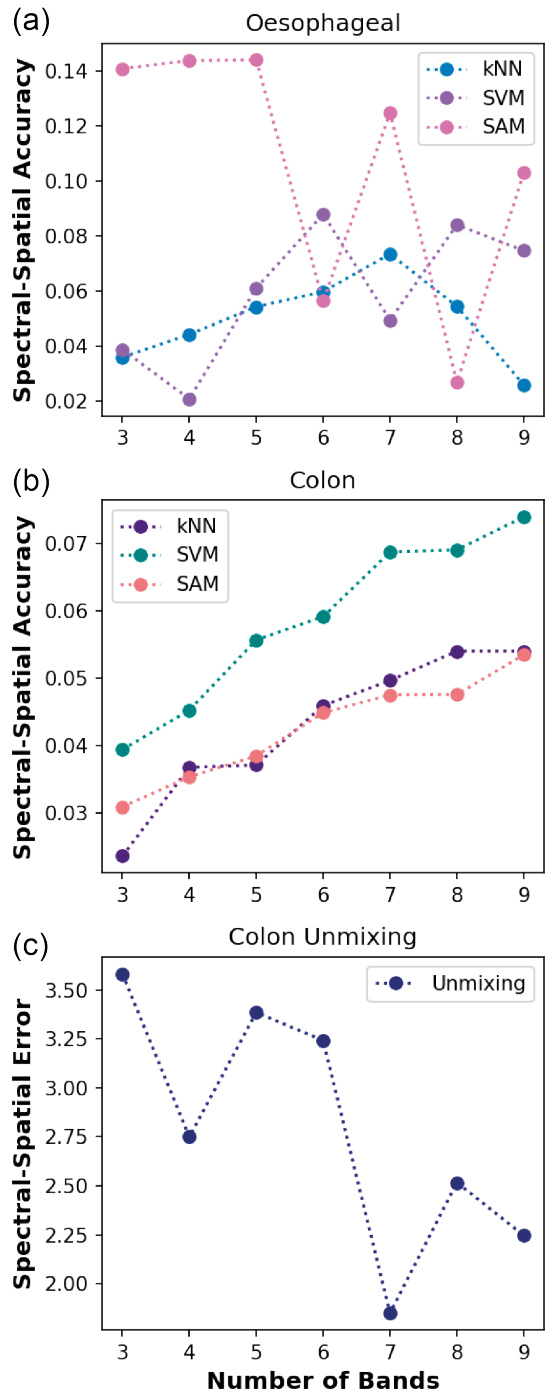

**Supplementary Figure 4.** The difference between the spatial and spectral classification accuracies for **(a)** oesophageal and **(b)** colon tissue is shown. The equivalent for unmixing of the colon is shown **(c)** but the difference between the spectral and spatial classification accuracies is shown since a lower unmixing error is desired. Note that a higher classification accuracy is desired **(a,b)** but in the case of unmixing error **(c)** a lower value indicates better performance and less error.

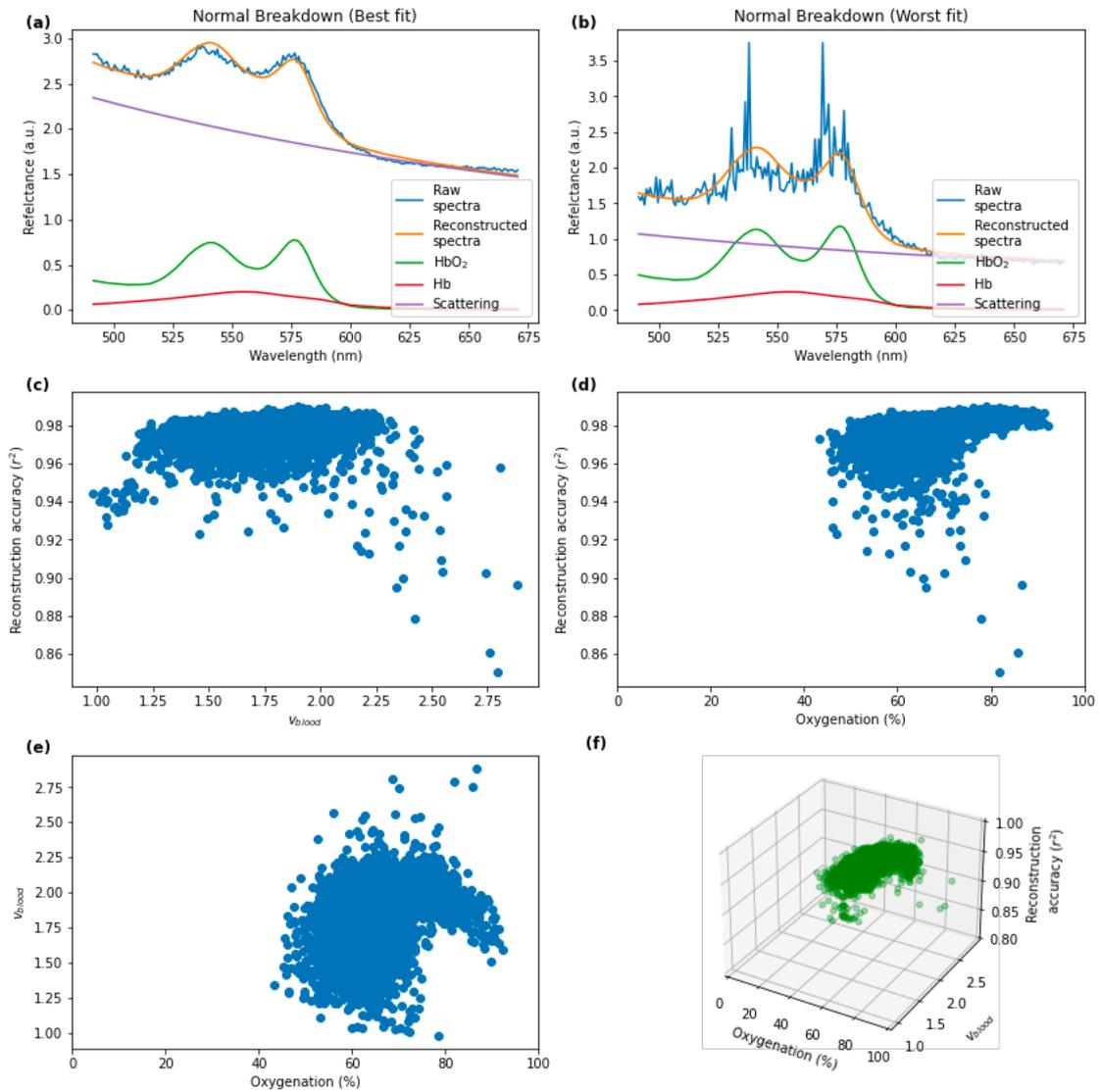

**Supplementary Figure 5: Normal tissue spectral fitting** (a) Raw and reconstructed spectra for the sample for the best fit example. The associated $HbO_2$, Hb, and scattering components that make up the reconstruction are shown. Please note the offset has been removed from both raw and reconstructed data to enable appropriate comparison; this was done for all further plots of raw and reconstructed spectra. (b) Raw and reconstructed spectra for the sample with the worst fit. (c) Reconstruction accuracy compared to the amount of blood (sum of Hb and $HbO_2$ values) for all normal tissue. (d) Reconstruction accuracy compared for the calculated oxygenation. (e) The amount of blood and oxygenation are compared. (f) A 3D plot illustrating reconstruction accuracy, the amount of blood, and oxygenation of normal tissue spectra.

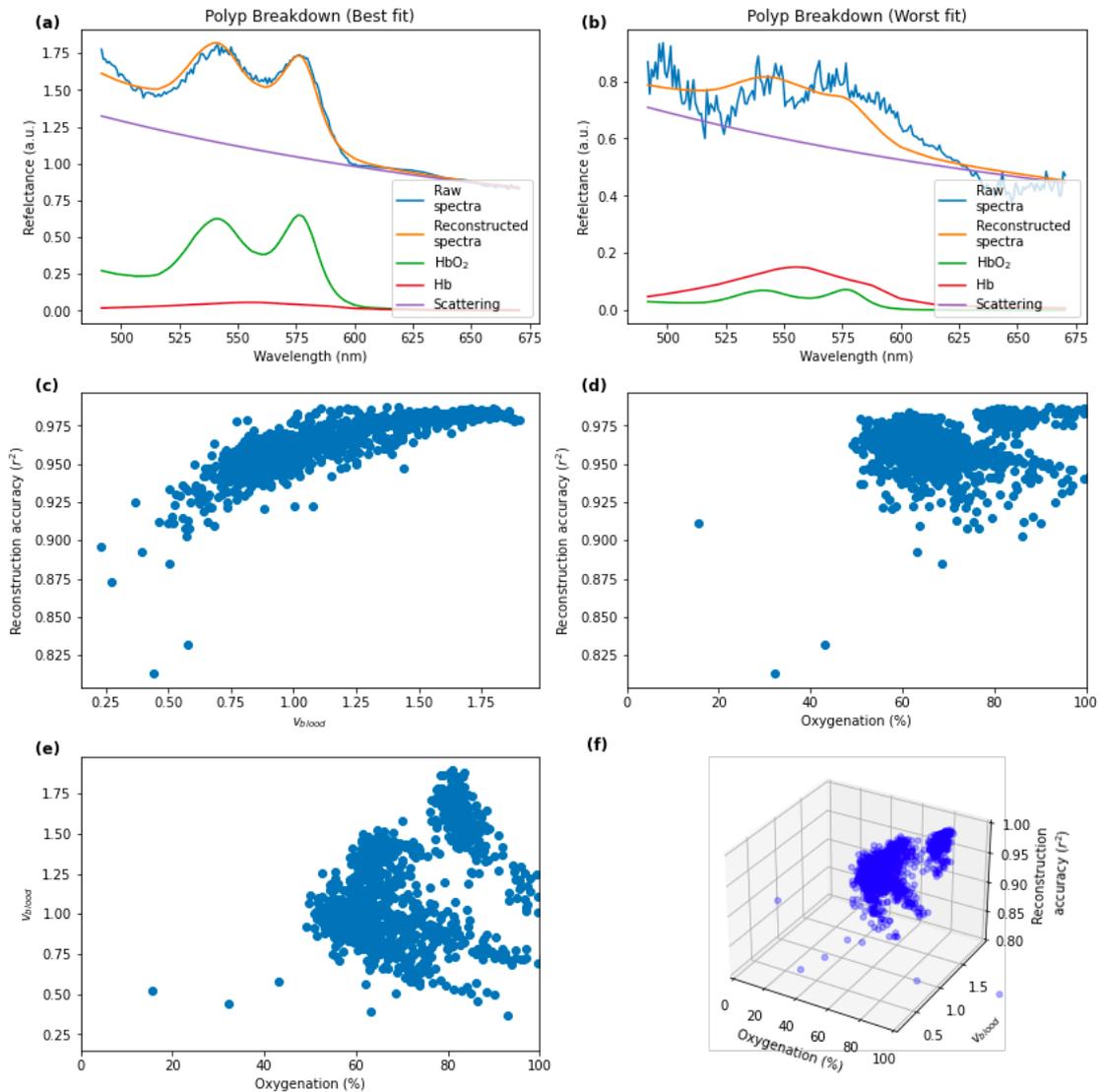

**Supplementary Figure 6: Polyp tissue spectral fitting** (a) Raw and reconstructed spectra for the sample for the best fit example. The associated $HbO_2$, Hb, and scattering components that make up the reconstruction are shown. (b) Raw and reconstructed spectra for the sample with the worst fit. (c) Reconstruction accuracy compared to the amount of blood (sum of Hb and $HbO_2$ values) for all normal tissue. (d) Reconstruction accuracy compared for the calculated oxygenation. (e) The amount of blood and oxygenation are compared. (f) A 3D plot illustrating reconstruction accuracy, the amount of blood, and oxygenation of normal tissue spectra.

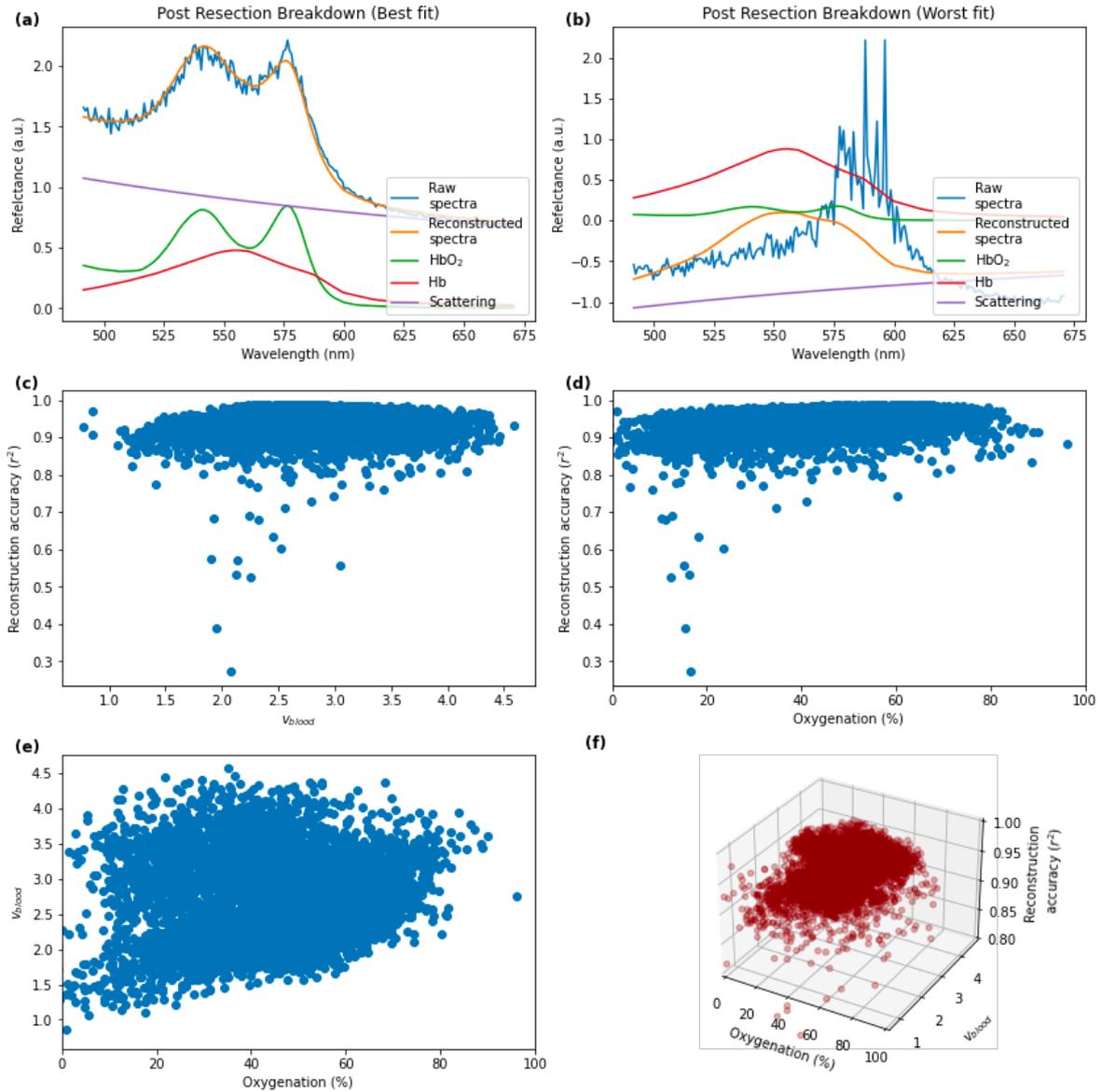

**Supplementary Figure 7: Post-resection tissue spectral fitting.** (a) Raw and reconstructed spectra for the sample for the best fit example. The associated $HbO_2$, Hb, and scattering components that make up the reconstruction are shown. (b) Raw and reconstructed spectra for the sample with the worst fit. (c) Reconstruction accuracy compared to the amount of blood (sum of Hb and $HbO_2$ values) for all normal tissue. (d) Reconstruction accuracy compared for the calculated oxygenation. (e) The amount of blood and oxygenation are compared. (f) A 3D plot illustrating reconstruction accuracy, the amount of blood, and oxygenation of normal tissue spectra.

**Supplementary Table 1**: **kNN Classification of oesophageal and colon data set.** The optimised MSFA for the oesophageal dataset, and the resulting classification of the synthetic hypercube are shown in the second and third column, respectively. The optimised MSFA for the colon dataset, and the resulting classification of the synthetic hypercube are shown in the fourth and fifth column, respectively.

| No of Bands | Oesophageal MSFA | Oesophageal Classification | Colon MSFA | Colon Classification |
|---|---|---|---|---|
| 3 | 475 (24) / 617 (16) / 617 (16) / 573 (22) | | 555 (20) / 687 (20) / 687 (20) / 601 (20) | |
| 4 | 475 (18) / 703 (16) / 565 (18) / 619 (24) | | 495 (20) / 673 (20) / 587 (20) / 685 (20) | |

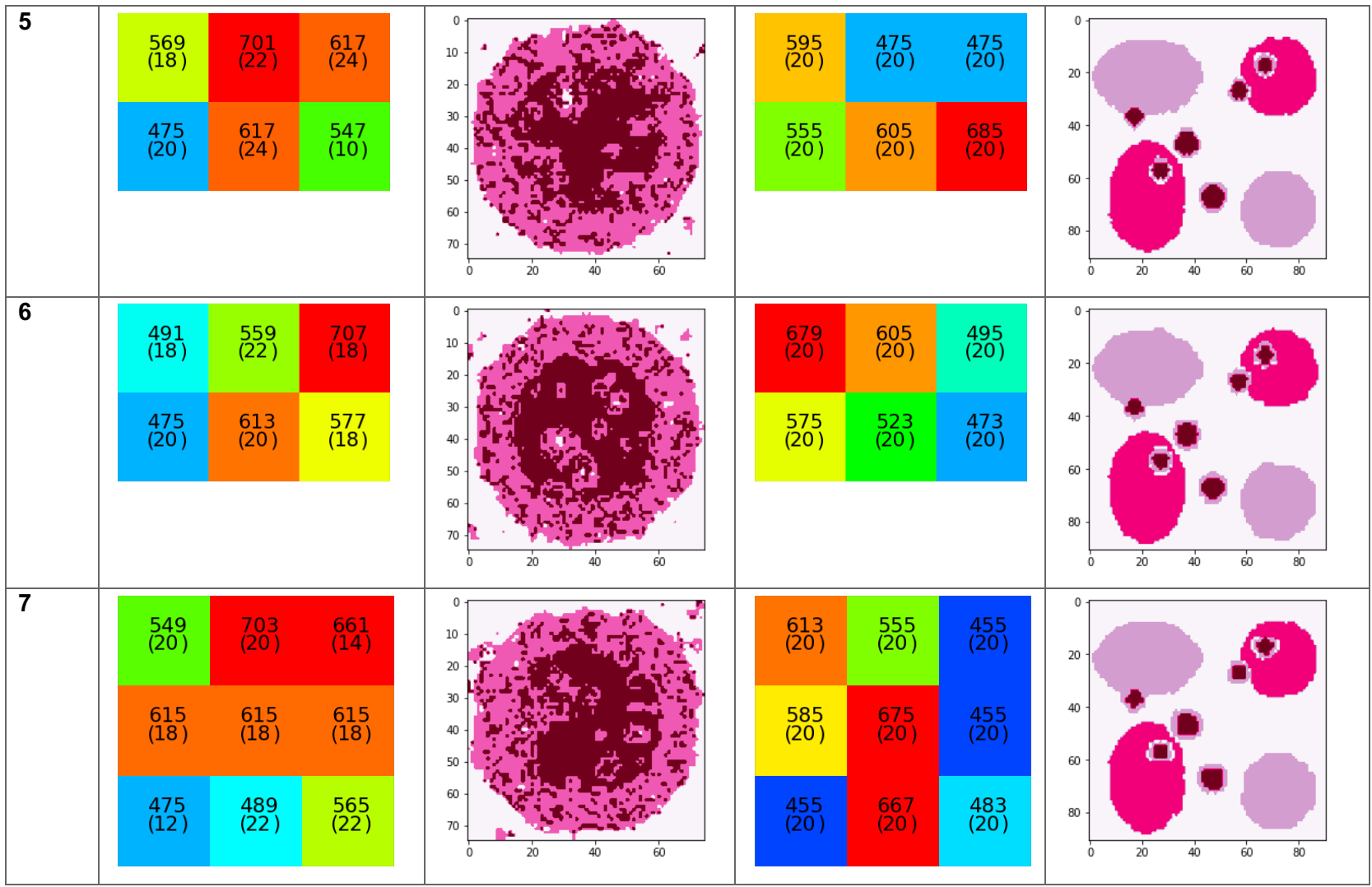

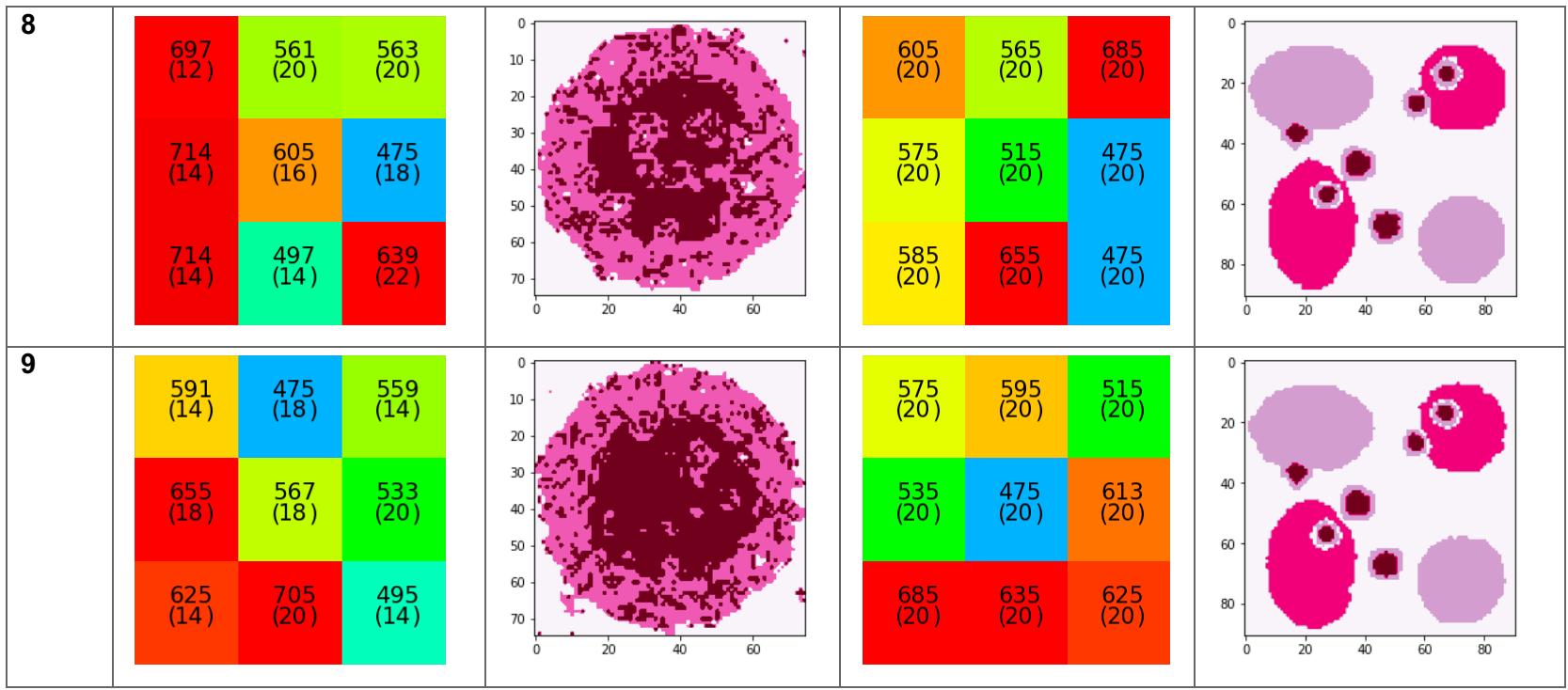

**Supplementary Table 2**: **SVM Classification of oesophageal and colon data set.** The optimised MSFA for the oesophageal dataset, and the resulting classification of the synthetic hypercube are shown in the second and third column, respectively. The optimised MSFA for the colon dataset, and the resulting classification of the synthetic hypercube are shown in the fourth and fifth column, respectively.

| No of Bands | Oesophageal MSFA | Oesophageal Classification | Colon MSFA | Colon Classification |
|---|---|---|---|---|
| 3 | 545 (16) / 565 (26) / 601 (20) / 601 (20) | 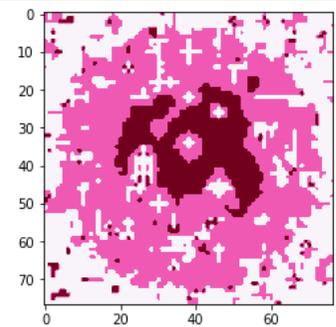 | 603 (20) / 685 (20) / 555 (20) / 685 (20) | 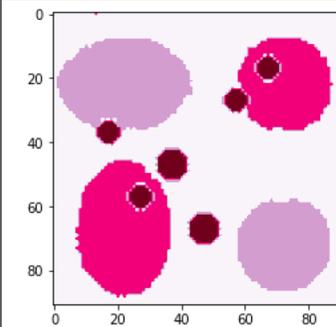 |

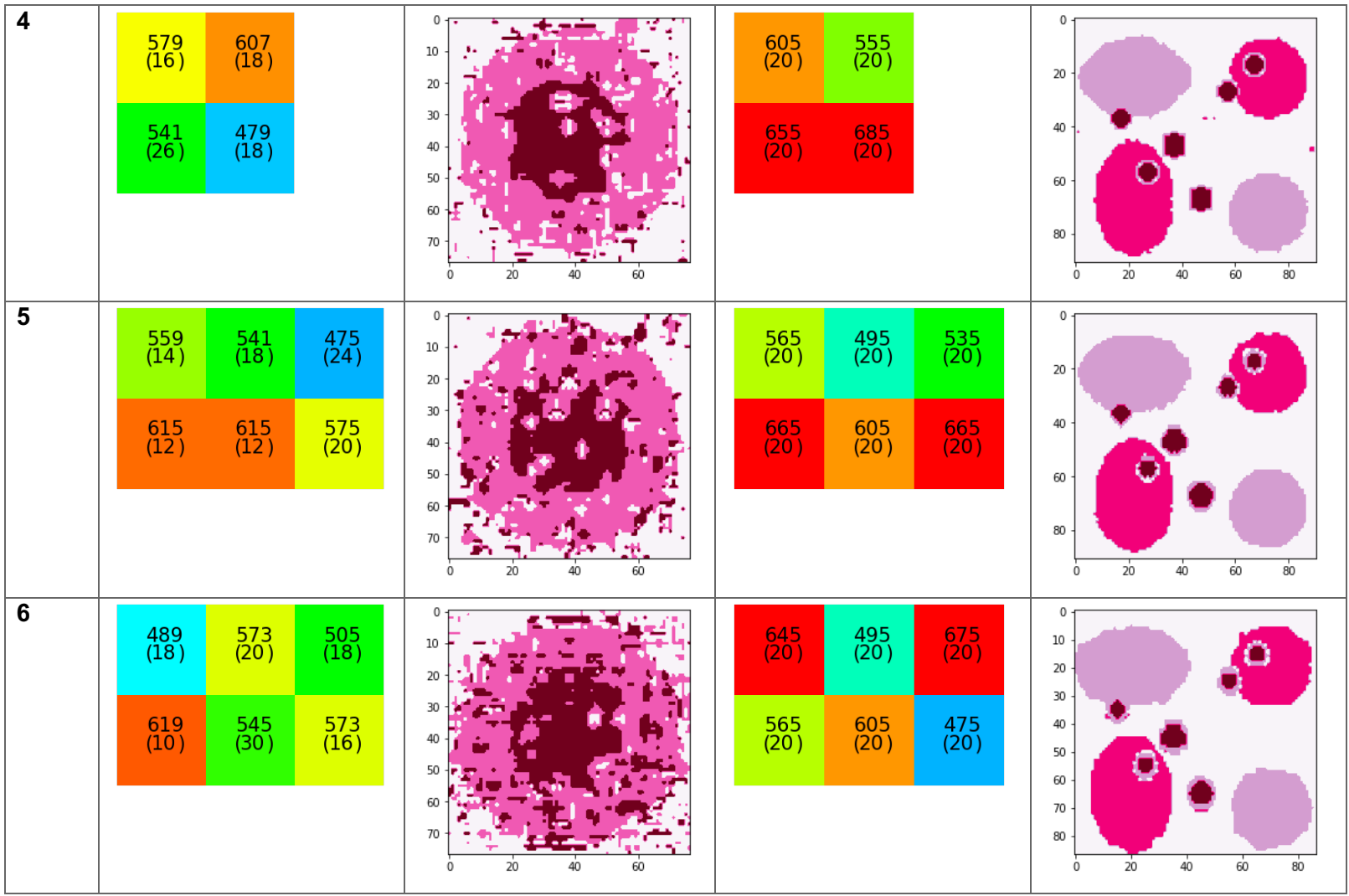

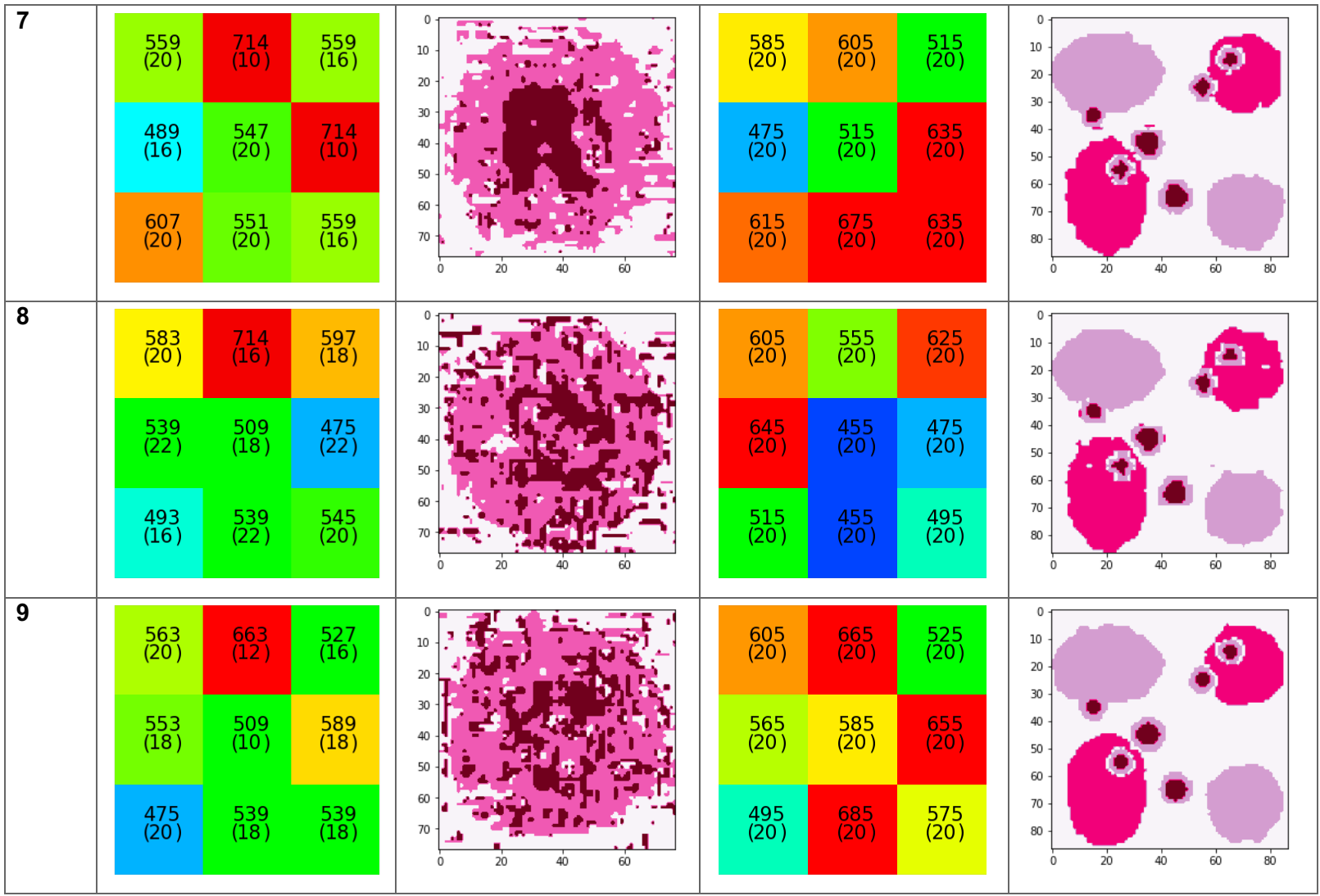

**Supplementary Table 3**: **SAM Classification of oesophageal and colon data set.** The optimised MSFA for the oesophageal dataset, and the resulting classification of the synthetic hypercube are shown in the second and third column, respectively. The optimised MSFA for the colon dataset, and the resulting classification of the synthetic hypercube are shown in the fourth and fifth column, respectively.

| No of Bands | Oesophageal MSFA | Oesophageal Classification | Colon MSFA | Colon Classification |
|---|---|---|---|---|

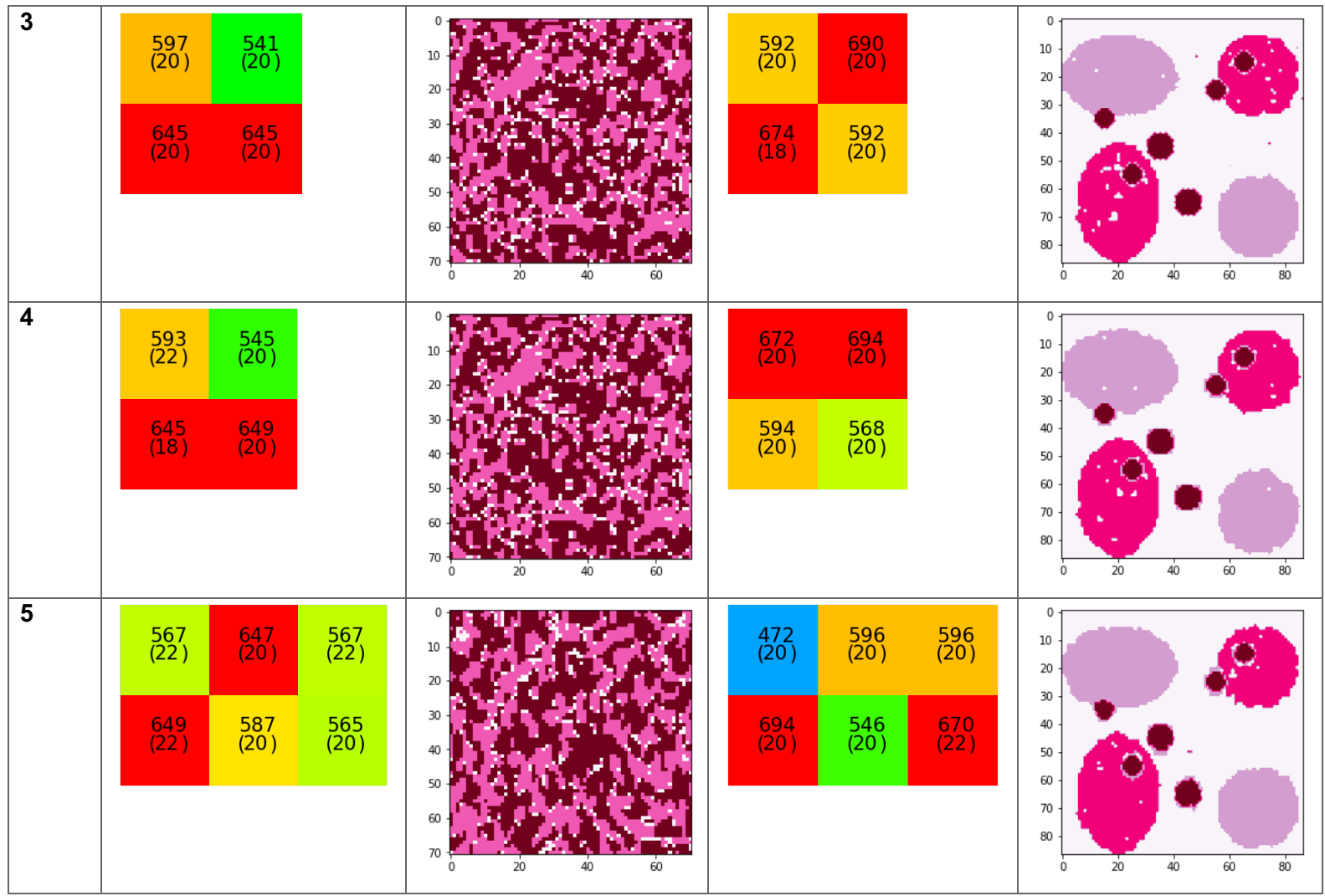

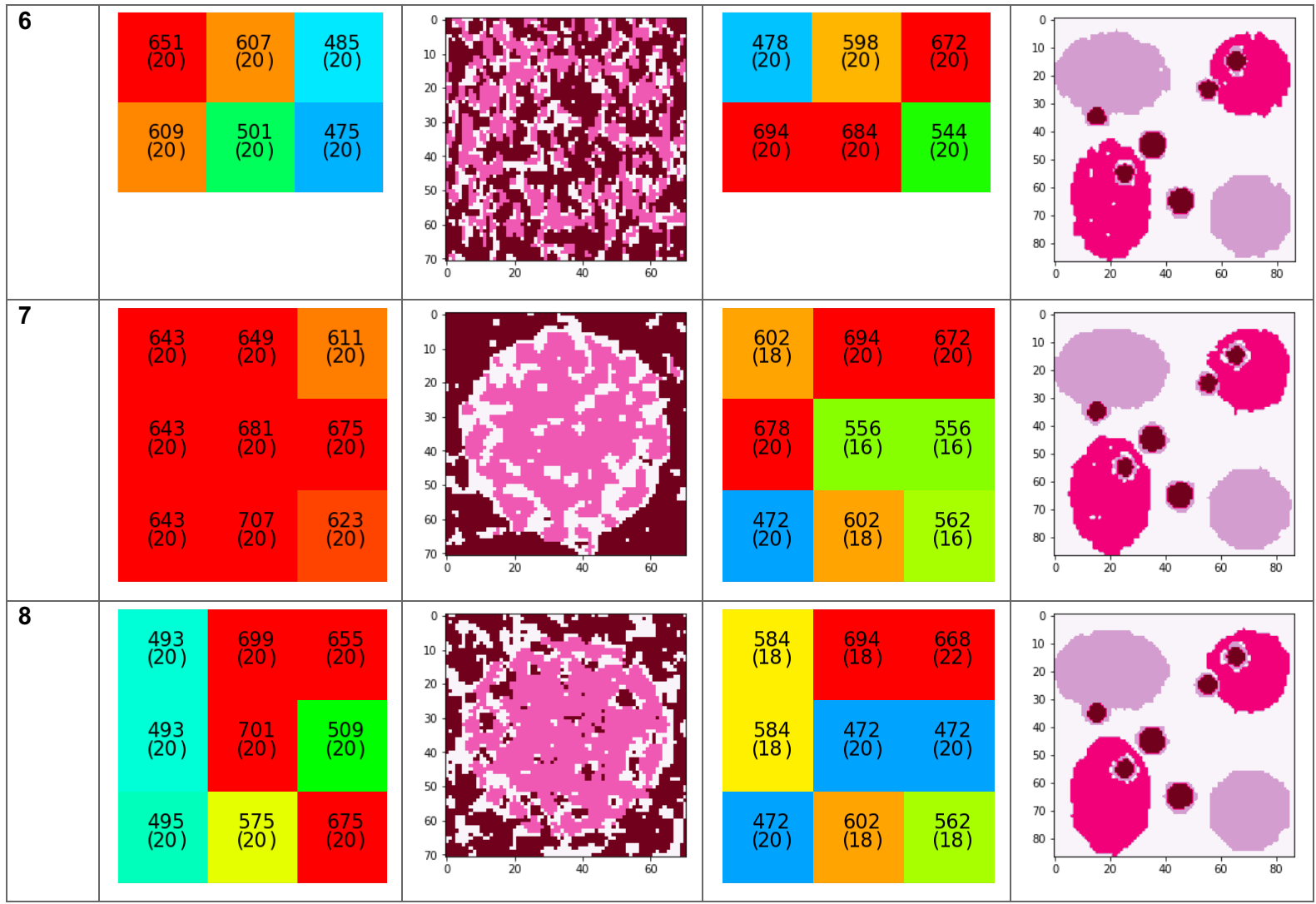

| 9 | 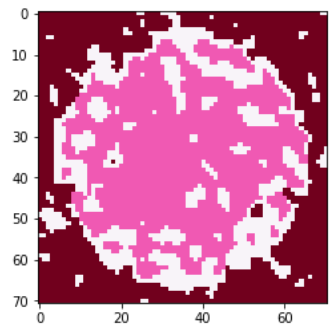 | | 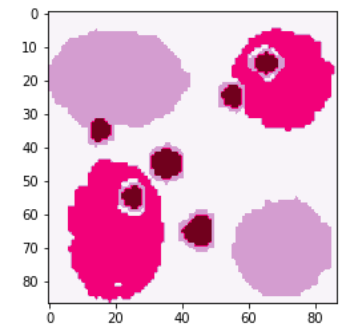 | |

**Supplementary Table 4: Normal tissue** Best and worst fit calculated values: $r^2$, $HbO_2$, Hb, water, scattering, and offset for normal tissue.

|           | $r^2$  | $HbO_2$ | Hb     | Water   | Scattering | Offset  |
|-----------|--------|---------|--------|---------|------------|---------|
| Best fit  | 0.9886 | 1.6863  | 0.4970 | 14.9042 | 20.8456    | 16.9042 |
| Worst fit | 0.8184 | 1.4141  | 0.6235 | 11.4896 | 14.4088    | 17.8684 |

**Supplementary Table 5: Polyp tissue** Best and worst fit calculated values: $r^2$, $HbO_2$, Hb, water, scattering, and offset for polyp tissue.

|           | $r^2$  | $HbO_2$ | Hb     | Water   | Scattering | Offset  |
|-----------|--------|---------|--------|---------|------------|---------|
| Best fit  | 0.9857 | 1.5593  | 0.2969 | 12.9823 | 13.7589    | 17.8368 |
| Worst fit | 0.5415 | 0.5091  | 0.9167 | 36.7926 | 21.5530    | 16.1552 |

**Supplementary Table 6: Post Resection Tissue.** Best and worst fit calculated values $r^2$, $HbO_2$, Hb, water, scattering, and offset for post-resection tissue unmixing.

|           | $r^2$  | $HbO_2$ | Hb      | Water    | Scattering | Offset  |
|-----------|--------|---------|---------|----------|------------|---------|
| Best fit  | 0.9902 | 1.4098  | 1.2150  | -0.2093  | 6.8400     | 19.1881 |
| Worst fit | 0.4763 | 1.3215  | -0.9968 | -46.5925 | -25.0397   | 25.1510 |

**Supplementary Table 7: Hb and HbO₂ unmixing of the colon data set.** The optimised MSFA for the colon dataset and the resulting estimates of sO$_2$ for the synthetic hypercube are shown in the second and third columns, respectively.

| No of Bands | Optimised MSFA based on unmixing | sO$_2$ |
|---|---|---|
| 3 | 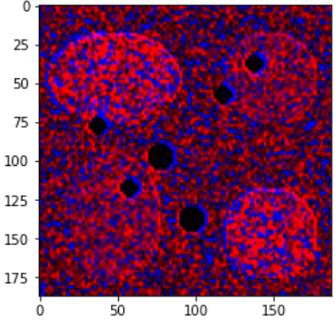 562 (14), 574 (12), 586 (30), 586 (30) | 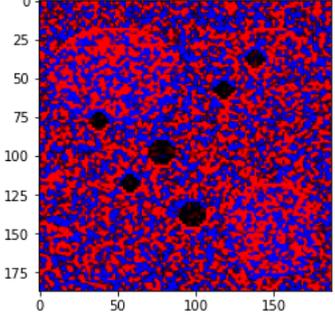 |
| 4 | 618 (30), 558 (30), 510 (10), 558 (18) | |
| 5 | 564 (14), 492 (10), 602 (18), 576 (10), 576 (10), 630 (12) | 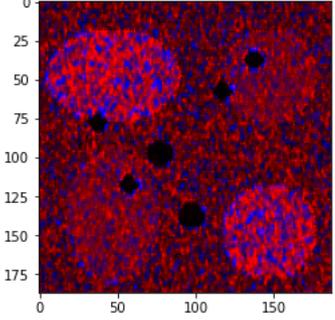 |
| 6 | 560 (16), 496 (10), 600 (22), 540 (12), 578 (16), 630 (16) | 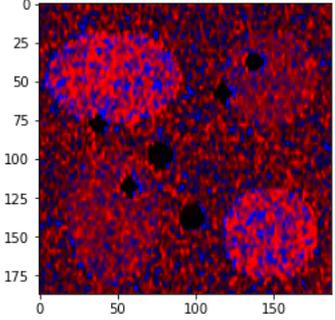 |

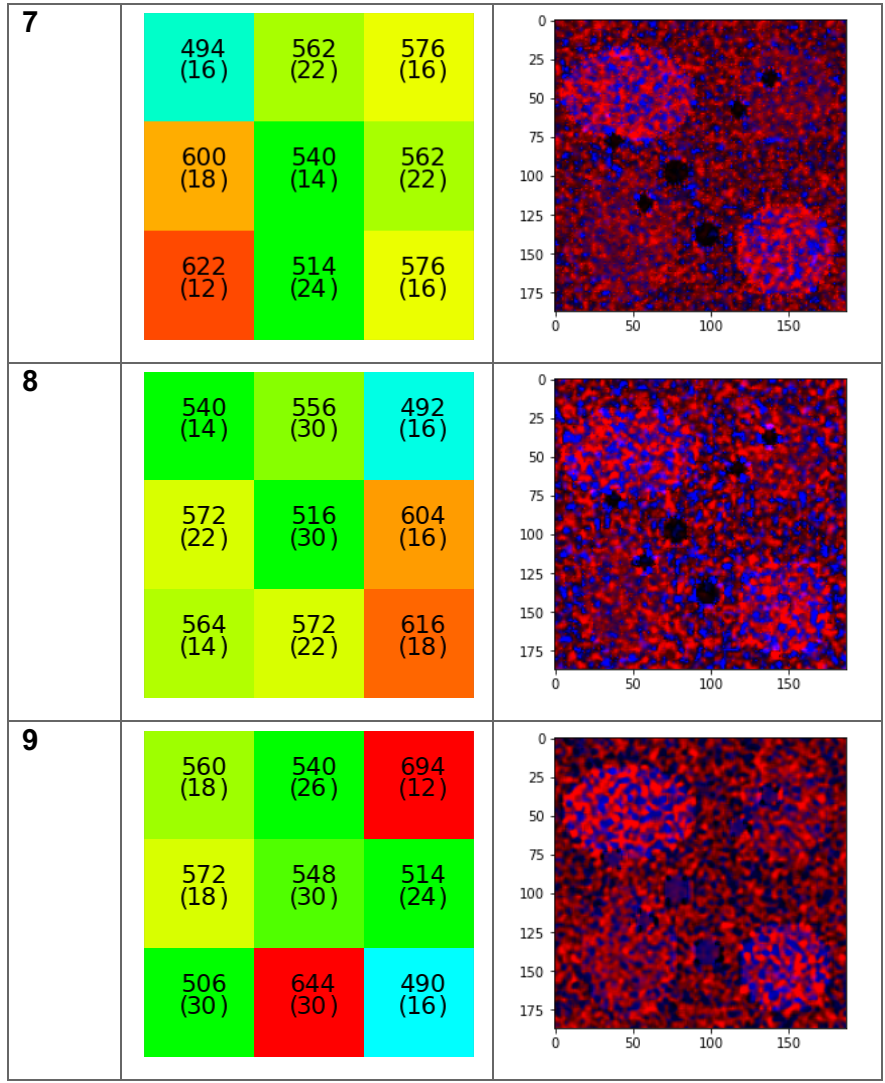

**References**

1. Yoon, J.; Joseph, J.; Waterhouse, D.J.; Borzy, C.; Siemens, K.; Diamond, S.; Tsikitis, V.L.; Bohndiek, S.E. First Experience in Clinical Application of Hyperspectral Endoscopy for Evaluation of Colonic Polyps. *J. Biophotonics* **2021**, 1–9, doi:10.1002/jbio.202100078.

2. Bosschaart, N.; Edelman, G.J.; Aalders, M.C.G.; Van Leeuwen, T.G.; Faber, D.J. A Literature Review and Novel Theoretical Approach on the Optical Properties of Whole Blood. *Lasers Med. Sci.* **2014**, *29*, 453–479, doi:10.1007/s10103-013-1446-7.

3. Prahl, S.A. Tabulated Molar Extinction Coefficient for Hemoglobin in Water Available online: http://omlc.ogi.edu/spectra/hemoglobin/summary.html.